\documentclass{aa}

\usepackage{natbib}
\bibpunct{(}{)}{;}{a}{}{,} 
\usepackage{graphics}   
\usepackage{graphicx}   
\usepackage{supertabular}
\usepackage{hyperref}   

\hypersetup{
    bookmarks=true,         
    unicode=false,          
    colorlinks=true,       
    linkcolor=blue,          
    citecolor=blue,        
    filecolor=blue,      
    urlcolor=blue           
}
\begin{document}

\title{A new infrared Fabry-P\'erot-based radial-velocity-reference module for the SPIRou radial-velocity spectrograph.\\[4ex]}

\author{Federica Cersullo\inst{1}, Fran\c{c}ois Wildi \inst{1}, Bruno Chazelas  \inst{1}, Francesco Pepe\inst{1}, 
}

\institute{Geneva Observatory, University of Geneva, Maillettes 51, CH-1290 Sauverny, Switzerland
}
\authorrunning{Cersullo et al.}
\titlerunning{A new infrared Fabry-P\'erot.}

\date{Received 28-10-2016/
Accepted 30-01-2017}
\abstract  {The field of exoplanet research is moving towards the detection and characterization of habitable planets. These exo-Earths can be easily found around low-mass stars by using either photometric transit or radial-velocity (RV) techniques. In the latter case the gain is twofold because the signal induced by the planet of a given mass is higher due to the more favourable planet-star mass ratio and because the habitable zone lies closer to the star. However, late-type stars emit mainly in the infrared (IR) wavelength range, which calls for IR instruments.}
{SPIRou is a stable RV IR spectrograph addressing these ambitious scientific objectives. As with any other spectrograph, calibration and drift monitoring is fundamental to achieve high precision. However, the IR domain suffers from a lack of suitable reference spectral sources. Our goal was to build, test and finally operate a Fabry-P\'erot-based RV-reference module able to provide the needed spectral information over the full wavelength range of SPIRou.}
{We adapted the existing HARPS Fabry-P\'erot calibrator for operation in the IR domain. After manufacturing and assembly, we characterized the FP RV-module in the laboratory before delivering it to the SPIRou integration site. In particular, we measured finesse, transmittance, and spectral flux of the system.} 
{The measured finesse value of $F=12.8$ corresponds perfectly to the theoretical value. The total transmittance at peak is of the order of 0.5\%, mainly limited by fibre-connectors and interfaces. Nevertheless, the provided flux is in line with the the requirements set by the SPIRou instrument. Although we could test the stability of the system, we estimated it by comparing the SPIRou Fabry-P\'erot with the already operating HARPS system and demonstrated a stability of better than 1\,m\,s$^{-1}$ during a night.} 
{Once installed on SPIRou, we will test the full spectral characteristics and stability of the RV-reference module. The goal will be to prove that the line position and shape stability of all lines is better than 0.3\,m\,s$^{-1}$ between two calibration sequences (typically 24 hours), such that the RV-reference module can be used to monitor instrumental drifts. In principle, the system is also intrinsically stable over longer time scales such that it can also be used for calibration purposes.}

\maketitle
\authorrunning{Cersullo et al.}
\titlerunning{A new infrared Fabry-P\'erot.}

\section{Introduction}
\indent SPIRou is a near-infrared (NIR) spectro-polarimeter and velocimeter that is soon to be offered as a new-generation instrument on the Canada-France-Hawaii Telescope (CFHT) \citep{Delfosse13}. Its main goals are a) the search and characterization of habitable exo-planets orbiting low-mass $\&$ very-low mass stars by using high-precision radial-velocity (RV) measurements and b) the study of the impact of magnetic fields in star and planet formation by means of spectro-polarimetry. SPIRou is presently being integrated in Toulouse and will be transferred to the CFHT in 2017. Operations are planned to start end of 2017.\\
The radial-velocity method is employed to determine the Doppler shifts of absorption lines in the host star's spectrum. M dwarfs present the opportunity to discover potential habitable planets due the larger radial-velocity signature induced by the planet on the lower-mass star \citep{Udry2007}. Also, the habitable zone (HZ) of an M star lies much closer in than that of a solar-type star, which again leads to a higher radial-velocity signal compared to a solar-type star. However, because M-dwarfs emit most of their flux in the infrared wavelength domain \citep{Reiners2010}, a radial-velocity instrument has to operate at infrared wavelengths, to make best use of the stellar light, .\\
The wavelength range, resolution and the design of SPIRou, as well as its stability, have been optimized to search for habitable planets around M stars. The expected RV amplitude induced by a two Earth-mass planet in the HZ of an M5 dwarf, for instance, is of the order of \,1.7\,m\,s$^{-1}$ \citep{Barnes11}. To achieve sufficiently significant detection, a measurement precision, and thus instrumental precision, of the order or better than 1\,m\,s$^{-1}$ is required. It is well known that such a precision can only be achieved if both the spectral calibration and the instrumental drift are controlled to the same level. Both wavelength calibration and drift measurements, make use of spectral references that fix the necessary wavelength scale. The RV-observable corresponds to the measurement of the central wavelength of a spectral line that has to be referred to a known laboratory standard. Unlike the visible case, for which multiple wavelength calibration sources exist, such as spectral lamps, iodine cells, and lately also laser frequency combs (LFC) \citep{Ycas2012, Schwab2015} and Fabry-P\'erots \'etalons \citep{Reiners2014}, in the infrared wavelength domain we are lacking suitable references sources.\\
The precision limitations in the calibration of radial velocity measurements play a key-role, especially when targeting the ambitious cases listed above. For these kind of investigations, an ultra-stable, broadband, bright, flexible and reliable wavelength calibrator is required.\\
To ensure optimal accuracy and precision, the spectrum of the ideal calibrator must have the following main characteristics:
\begin{itemize}
\item Many unresolved narrow lines
\item Coverage of the full spectral range
\item Known (or calibrated) line position
\item Line stability
\item No blending
\item Homogeneous peak flux
\end{itemize}

In recent years, we developed a Fabry-P\'erot calibration system with the aforementioned characteristics \citep{Wildi2010} for the visible wavelength range. The first system, currently in use on the HARPS spectrograph, has performed well and has completely replaced the ThAr spectral lamp for instrumental drift measurement purposes. However, its lens-based optical design was not compatible with an infrared version. Therefore, we redesigned the system using an all-reflective solution. Two systems of the new design have been manufactured and are  currently in operation on the HARPS-N@TNG and CORALIE@EULER spectrographs. Both systems have demonstrated stability of better than 1\,m\,s$^{-1}$ during an observing night that allowed us to use them systematically for the simultaneous reference measurement \citep{Pepe2004}. Based on this second design, we have manufactured and tested a similar system for SPIRou. The only element that needed to be adapted was the \'etalon itself in terms of material and coatings.\\
In the next section we will describe the basic working principles of the Fabry-P\'erot based RV-reference module for SPIRou. In Sect. \ref{ref:three} we will describe the opto-mechanical design and the rationale that led us to the final parameter choice. In Sect. \ref{ref:four}, we present the laboratory measurement set-up  and the results obtained. In the final sections, we will also describe some of the particular aspects of RV precision and we will provide an outlook on possible future developments.

\section{Theory of the plano-plano Fabry-P\'erot \'etalons}\label{ref:theory}
A plano-plano Fabry-P\'erot \'etalon (FP) consists of two flat parallel mirrors separated by an optical gap $d$. The inner surface of each mirror is coated with a partially reflecting coating and the outer surface has an anti-reflection coating. The medium between the mirrors is transparent and has a refractive index of $n$.\\
\begin{figure}[htbp]
 \centering
\includegraphics[scale=0.50]{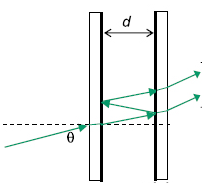}
\caption{Concept of the plano-plano FP and sketch of the interfering light beams}
\label{fig:1}
\end{figure}

The FP is illuminated with a parallel light beam of divergence $\Phi$. The light is partially reflected and transmitted by both mirrors producing interference in both reflection and transmission as shown in Figure \ref{fig:1} (excerpt from Field guide to spectroscopy of \citet{ball2006}). The transmission function of the \'etalon, taking into account the losses due to unavoidable absorption and scattering, is given by:.
\begin{equation}
T=\frac{I_{trans}}{I_{inc}}=\left(1-\frac{A}{1-R}\right)^{2} \frac{1}{1+\frac{4R}{(1-R)^{2}}sin^{2}\frac{\delta}{2}}
\label{equa:transtot}
\end{equation}
where $R$ is the reflectivity the mirror, $A$ the absorption and scattering losses per reflection, and $\delta = 4 \pi n d cos \theta /\lambda$ the phase difference between successive beams. Defining the reflective finesse coefficient as $F_R=\frac{4R}{(1-R)^{2}}$ and assuming $A<<1-R$, the Eq.~\ref{equa:transtot} is simplified to:
\begin{equation}
T(\lambda)=\frac{1}{1+(2F_{R}/\pi)^{2}\,sin^{2}\frac{\delta(\lambda)}{2}}
\label{equa:transsempl}
\end{equation}
This function reaches a transmission maximum every time the phase $\frac{\delta}{2}$ is an integral multiple of $\pi$, namely,
 \begin{equation}
 \label{equa:FP}
 2ndcos\theta = m \lambda
 \end{equation}
The distance between two successive maxima, also called the free spectral range (FSR), is defined by:
\begin{equation}
FSR=\frac{\lambda}{m}=\frac{\lambda^{2}}{2ndcos\theta}
\label{equa:FSR}
\end{equation}

\begin{figure}[htbp]
\centering
\includegraphics[width=10cm,height=7cm]{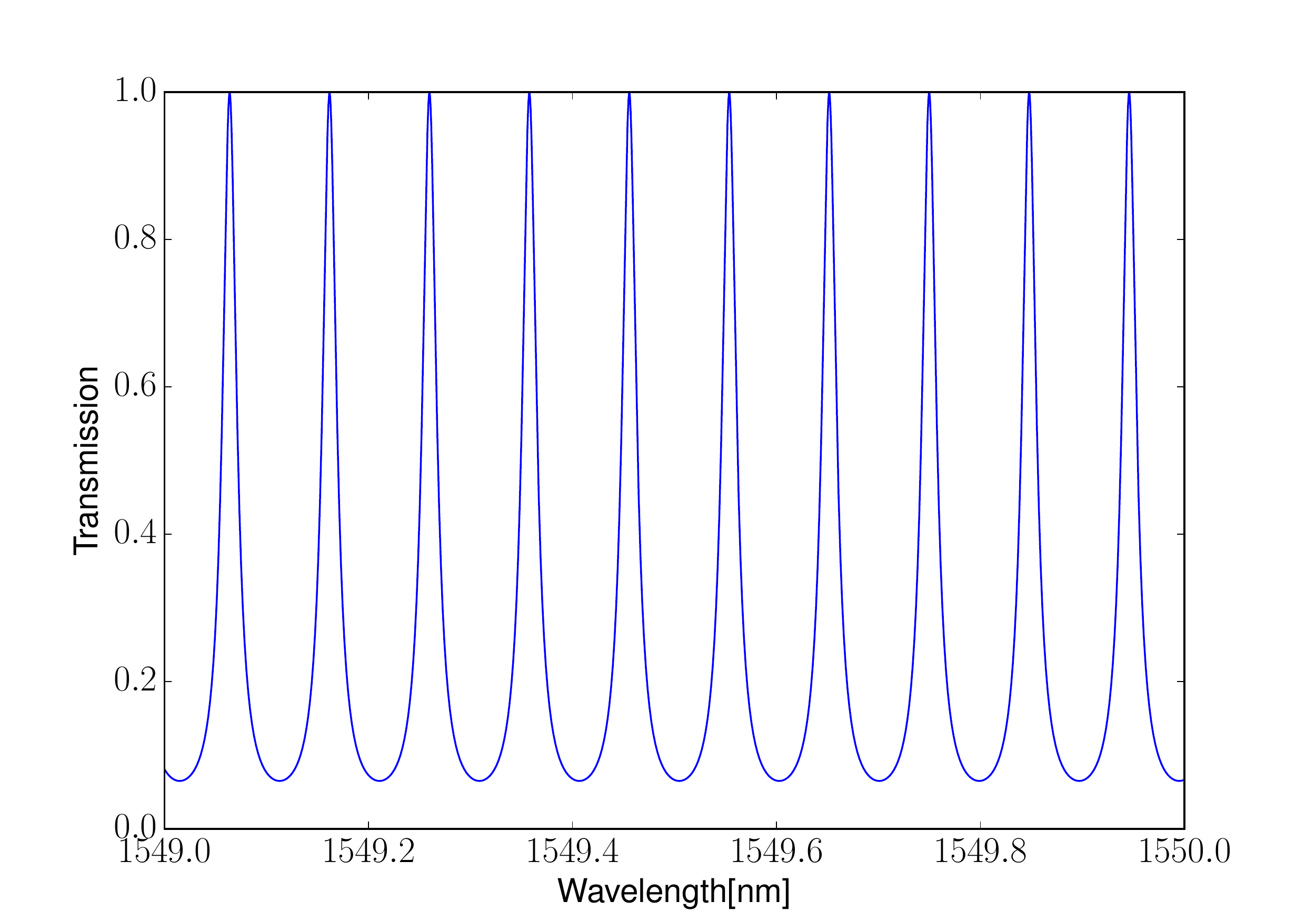}
\caption{Simulated curve for our Fabry-P\'erot passband.}
\label{fig:2}
\end{figure}

Figure \ref{fig:2} shows the theoretical transmission function of a Fabry-P\'erot \'etalon with finesse $F_R=13$ at $1540\, nm$ wavelength and FSR of $\sim 0.09\, nm$. In this example we chose normal incidence ($\theta=0$) and vacuum operation ($n=1$). Transmission maxima are then reached for each wavelength $\lambda_m=2d/m$ and the free spectral range simplifies to $FSR=\lambda^2/2d$.\\

The finesse $F$ of a Fabry-P\'erot \'etalon is defined as the ratio of the FSR to the full-width at half maximum (FWHM) of the transmission peak $F=\frac{FSR}{FWHM}$. The functional analysis leads to $FWHM=\frac{\lambda}{mF_R}$, leading to $F=F_R$.\\
We  previously assumed a perfect \'etalon. In reality, any \'etalon will show errors such as surface errors and non-parallelism of the mirror. In order to take into account all these contributions, we can replace the reflective finesse $F_R$ by an effective finesse $F_{e}$ of the \'etalon, which can be expressed as:

\begin{equation}
\frac{1}{F^{2}_{e}}=\frac{1}{F^{2}_{R}}+\frac{1}{F^{2}_{D}}+\frac{1}{F^{2}_{P}}+\frac{1}{F^{2}_{\Phi}}
\label{equa:eff}
\end{equation}
where $F_{R}$ is the reflectivity finesse, $F_{D}$ is the defect finesse, and $F_{P}$ is the parallelism finesse. An approximation of the divergence finesse is represented by $F_{\Phi}$. This is produced by the finite dimension of the illuminating entrance slit, which in turn produces a beam divergence of the quasi-parallel beam through the \'etalon. In the simulations we will present in the following sections, we replaced the divergence finesse by an integration over the field positions of the slit, to compute a realistic transmission function of the \'etalon in the case of a large entrance slit.
 
Using the effective finesse and the definition of the resolving power $RP$, we computed:
\begin{equation}
RP:=\frac{\lambda}{\Delta \lambda}=\frac{\lambda}{FWHM} = \frac{\lambda}{\lambda/(mF_e)}= mF_e
\label{equa:res}
\end{equation}

\section{Infrared Fabry-P\'erot radial-velocity-reference module for SPIRou}\label{ref:three}
The Fabry-P\'erot RV-reference module for SPIRou has the purpose of providing a large number of ultra-stable lines in the wavelength range of the spectrograph, that allow to measure the instrumental drift during the night (or, in other terms, on the time scale between two calibrations). From the general characteristics of an ideal calibration source described in the introduction, we derived the requirements for the SPIRou RV-reference module as follows:

\begin{itemize}
\item{The calibration source must cover the full wavelength range of SPIRou, i.e. from 980\,nm to 2350\,nm.}
\item{The lines must not be resolved by the SPIRou spectrograph, since otherwise information is lost, respectively we do not take full advantage of the spectral resolution of the instrument.}
\item{The line separation must be minimum two and maximum four the full width at half maximum (FWHMs) of the spectrograph instrumental-profile (IP), in order to maximize the information content and avoid blends.}
\item{The line position and shape stability of all lines must be better than 0.3\,m\,s$^{-1}$ during a night and on the time scale in between two calibrations (typically 24 hours).}
\item{The lines must be equidistant and the position of all lines defined by a limited number of system parameters.}
\item{The relative intensity of any neighboring lines must be stable at $10\%$ over any time scale.}
\item{The dynamic range of line intensities must be smaller than a factor 2 over one echelle order and a within a factor five over the full spectrum.}
\end{itemize}

\subsection{Spectral characteristics}\label{par:spectral}
Assuming the aforementioned hypothesis, we were able to design the Fabry-P\'erot RV-reference module's spectral characteristics. First, we took into account the resolution of the spectrograph $R_{S}$. To optimize the spectral content, we need to make the line spacing as small as possible but also ensure that on the blue side, where it is minimal, it remains larger than $3\cdot FWHM$. This led us to the first condition: 
\begin{equation}
FSR_B \geq 3\cdot\frac{\lambda_B}{R_S}
\label{equa:espr}
\end{equation}

In the case of vacuum operation and perpendicular incidence, Eq.\,\ref{equa:FP} simplifies to
\begin{equation}
2 \cdot d=m \cdot \lambda.
\label{equa:FPsimple}
\end{equation}

Combining this equation with the inequality in Eq.\ref{equa:espr}, we obtained the condition for the \'etalon gap:
\begin{equation}
FSR_B=\lambda_B/m=\frac{\lambda_B^2}{2\cdot d}\geq 3\cdot\frac{\lambda_B}{R_S}
\end{equation}

If we set the gap to exactly match the minimum FSR, we obtained the value:

\begin{equation}
d=\frac{\lambda_B\cdot R_S}{6}.
\end{equation}

On the other hand, the lines of the \'etalon, at constant finesse, are larger on the red side of the spectrum. Therefore, we had to ensure that they remain narrower than the IP in this area of the spectrum. Our choice was to under-resolve the FP lines by a factor of two to three and not less, such that unnecessary flux loss is avoided. This choice drove the second condition:

\begin{equation}
FWHM \leq \frac{2}{3}\cdot\frac{\lambda_R}{R_S}
\label{equa:espr1}
\end{equation}

Using the definition of the finesse $F=FSR/FWHM$ and combining it with Eq.\ref{equa:espr1}, we obtained for the finesse:
\begin{equation}
F \geq \frac{2\cdot d\cdot 3 \cdot R_S}{m^2\cdot 2 \cdot \lambda_R}=\frac{3 \cdot d \cdot R_S \lambda_R^2}{4\cdot d^2 \lambda}=\frac{3 \cdot R_S \lambda_R}{4 \cdot d}
\end{equation}
Choosing the lowest possible finesse and by substituting the gap $d$, we obtained for the finesse:
\begin{equation}
F=\frac{3\cdot R_S\cdot \lambda_R}{4D}=\frac{9 \lambda_R}{2 \lambda_B}
\end{equation}
To summarize, the spectrograph's spectral range and resolution, together with the general requirements for the RV-module described in the previous section, defined the Fabry-P\'erot parameters summarized in Table\,\ref{tab:parameter}.\\

\begin{table}[h]
\centering 
\begin{tabular}{l l l}
\hline\hline
\textbf{Parameter} & \textbf{Variable} & \textbf{Value}\\ 
\hline
SPIRou start wavelength & $\lambda_B$ & 980\,nm \\ 
SPIRou end wavelength & $\lambda_R$ & 2350\,nm \\ 
SPIRou resolving power & $R_S$ & 75'000  \\ 
Étalon finesse & $F$ & 10.8  \\ 
Étalon gap  & $d$ & 12.250 \,mm  \\ 
\hline\hline
\end{tabular}
\caption{SPIRou's main parameters and derived \'etalon parameters}
\label{tab:parameter}
\end{table}

To achieve a reflectivity-limited effective finesse of about $F=11$, we distributed the various finesse components of Eq.\,\ref{equa:eff} as shown in Table \ref{tab:hresult}. As mentioned above, our starting choice was to determine the effective finesse by allocating a value of $F_R=13$ to the reflectivity finesse. The remainder of the budget was almost completely used up by the divergence finesse, to maximize the fibre diameter and minimize the physical size of the \'etalon. The remaining finesse values were set much higher to keep $F_{e}$ in budget.\\ 
\begin{table}[h]
\caption{Finesse budget and derived requirements}
\centering 
\begin{tabular}{l c c l}
\hline\hline 
\textbf{Finesse} & \textbf{Budget} & \textbf{Formula} & \textbf{Req.}\\
\hline 
Reflectivity & 13 & $F_R=\pi\sqrt(R)/(1-R)$ & $R=80\%$\\ 
Divergence & 28 & $F_{\Phi}=4\cdot\lambda/(\Phi^2 \cdot d)$ & $\Phi=2$\,mrad\\
Parallelism & 40 & $F_P=\lambda/(2\cdot\Delta)$ & $\Delta=4.5$\,nm  \\
Defect  & 40 & $F_D=\lambda/(\sqrt(2)\cdot\delta)$ & $\delta=6.7$\,nm \\
Effective & 10.88 & Eq.\,\ref{equa:eff} & \\
\hline\hline 
\end{tabular}
\label{tab:hresult}
\end{table}

\begin{figure}[htbp]
\centering
\includegraphics[width=9cm,height=5.5cm]{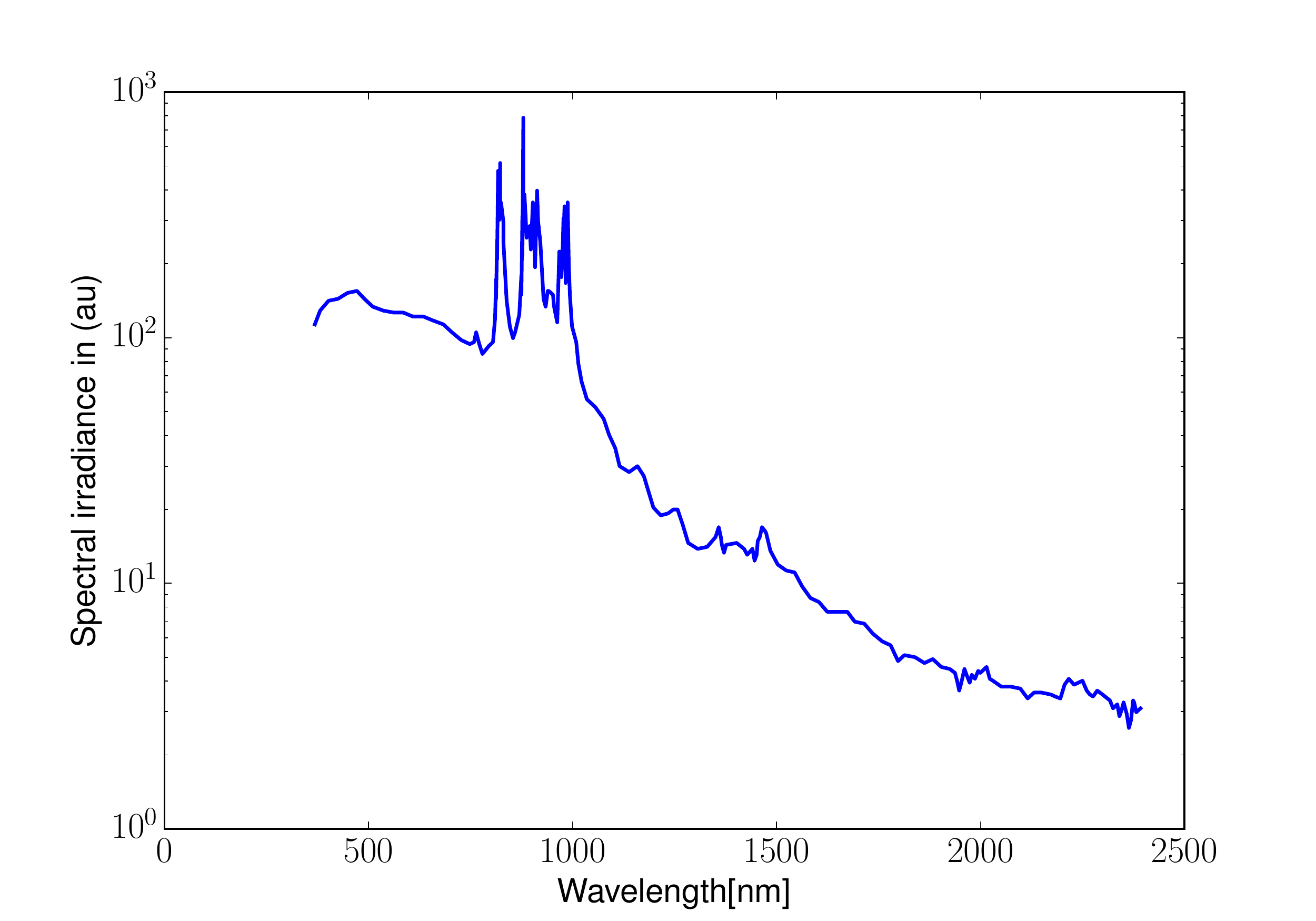}
\caption{Spectral power produced by the LDLS. The original plot, which is an excerpt from LDLS EQ-99 handbook, was manually digitized.}
\label{fig:3}
 \end{figure}

Once the finesse values  have been frozen, they define minimum requirements to the opto-mechanics of the \'etalon. The finite divergence forces a trade-off between fibre size, the focal length of the collimator, and clear aperture of the \'etalon. In principle, at fixed divergence value, the physical dimensions of the assembly are minimized when the smallest fibre is selected. However, this implies a loss of flux inversely proportional to the fibre size if an incoherent (extended) light source is used. To maximize the flux while still using an incoherent source, we decided to go for the source with the highest possible surface brightness. The choice fell on the Laser-Driven Light Source (LDLS) EQ-99 FC by Energetiq that is composed of a high-temperature plasma bulb $T=10000$\,K (Figure \ref{fig:3}, excerpt from LDLS EQ-99 handbook) heated by a high-power diode laser. The light output perfectly matches that of an ordinary step-index fibre of 200\,$\mu$m that can directly be used to feed our FP assembly. The numerical aperture $NA=0.20$ and the practical wish to keep the system compact, led us to fix the focal length of the collimator to 100\,mm and the clear aperture of the \'etalon to 40\,mm. The parallelism defect $\Delta\,<4.5\,nm$ translated into $10^{-6}$\,rad or 0.2\,arcsec. Finally, the value for the surface defects of $\delta=6.7$\,nm required the mirrors to be manufactured with at least $\lambda/50$ in wave-front error.
 
\subsection{Sensitivity of finesse and radial-velocity stability to alignment and fibre diameter}
Up to now we have considered the transmission function of the \'etalon to be given by Eq.\,\ref{equa:transsempl}. Defects, non-parallelism and finite divergence were taken into account by computing an effective finesse that was corrected for the various contributions. Although this approach is satisfactory for the defects and non-parallelism of the mirrors, it is incomplete for the finite divergence. The finite divergence (or fibre diameter) not only introduces a loss of finesse but also a shift in wavelength of the transmitted peaks. Furthermore, a non-uniform light distribution across the fibre exit produces a change in the shape of the transmitted peak. The resulting transmission function can be simply understood as the sum of the individual transmission functions from every field point (or, in other words, over each ray arising from the different points across the fibre tip) that will cross the \'etalon with slightly different incidence angles $\theta$.\\

We simulated the transmission function for different fibre diameters and as a function of the de-centring of the fibre from the optical axis, to understand how sensitive the finesse and the peak wavelength are as a function of these parameters. For the analysis, we needed to consider that the fibre is spatially extended and the deviation of each ray from the parallel beam is:
\begin{equation}
\theta=arctan\left(\frac{x}{f}\right)
\end{equation} 

where $x$ is the distance from the fibre centre and $f=100$\,mm is the focal length of the collimator (parabolic mirror). The angle $\theta$ by which the beam crosses the \'etalon depends on the position of the fibre of entrance.\\

To obtain the total transmittance spectrum, we need to compute the integral over the surface of the fibre. For the simulations we assume circular fibres and uniform illumination, and we perform the integral in polar coordinates. The total transmittance can then be written as:

\begin{equation}
T(\lambda)=\left(\int_{0}^{R} \rho d\rho \int_{0}^{2\pi} \frac{1}{1+(2\cdot F_{eff}/\pi)^2 \cdot sin^2(\delta/2)} d\varphi\right)/A
\label{equa:integral}
\end{equation}

where $R$ is the radius of the fibre and $\delta =4\pi n d cos\theta/\lambda$, as defined earlier. The integral is divided by the surface $A$ of the fibre tip to re-normalize the peak transmittance.\\

In case of a de-centring $\Delta$ of the fibre with respect to the optical axis of the system, the angle of incidence $\theta$ becomes:
\begin{equation}
\theta=arctan(\sqrt{(\rho cos\varphi+\Delta)^2+(\rho sin\varphi)^2}/f)
\end{equation}

where, without loss of generality, we have assumed the fibre to be de-centred along the x-axis only.\\

It is easily understood that, if the fibre has infinitely small diameter, all the the rays will cross the Fabry-P\'erot with $\theta\,=\,0$, such that we end up with the  known Airy function. For the general case of a finite fibre diameter, we performed a numerical integral to obtain the transmission as a function of the diameter of the fibre $D$ and $\Delta$.\\

\begin{figure}[htbp]
\centering
\includegraphics[width=9.5cm,height=7cm]{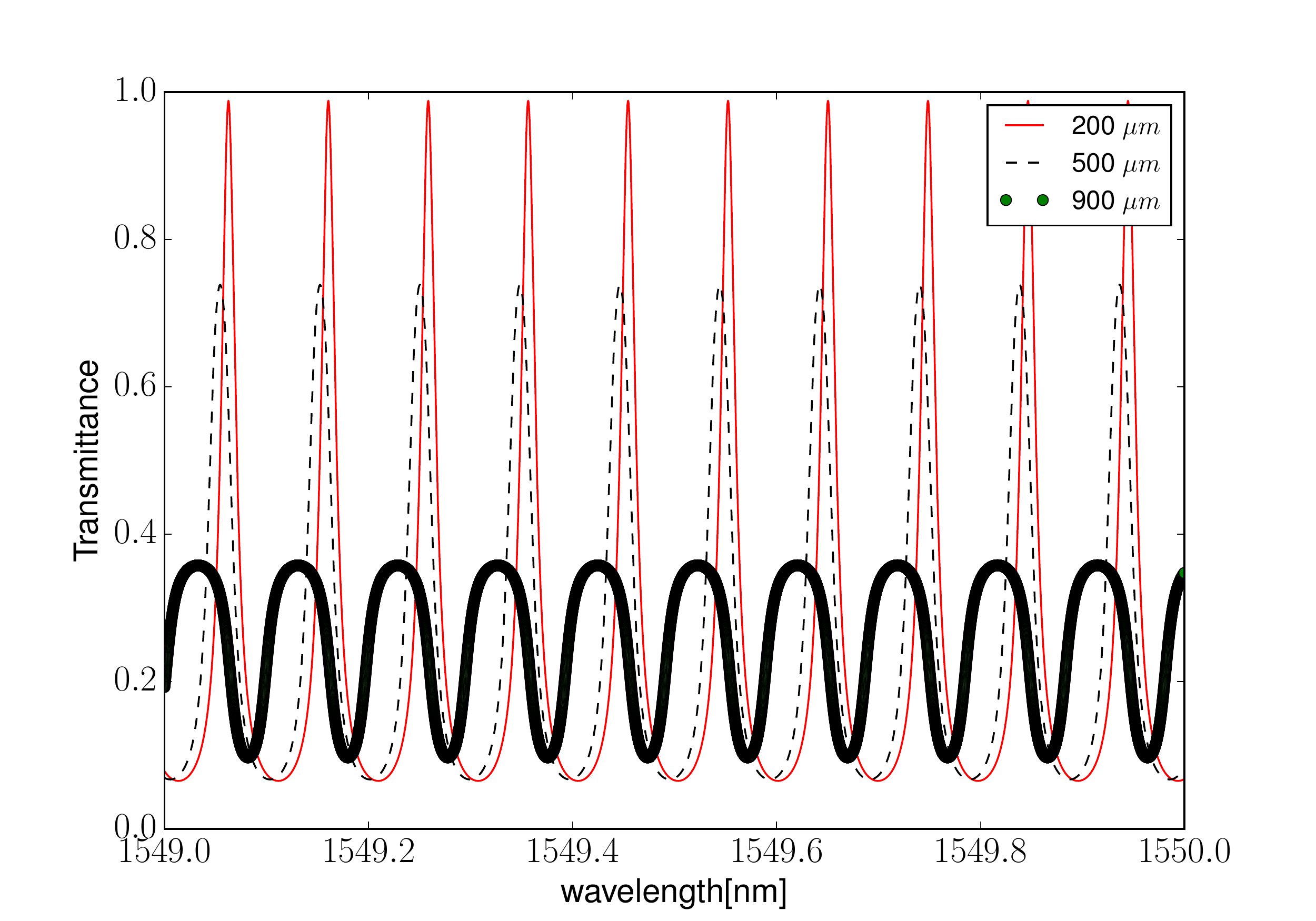}
\caption{Simulated transmittance of a Fabry-P\'erot for fibre diameters of 200, 500, and 900\,$\mu m$, assuming the fibre is centred on the optical axis and a focal length of the collimator $f=100$\,mm }
\label{fig:4}
\end{figure}

Figure \ref{fig:4} shows the transmittance as a function of fibre size for a system with focal length $f=100$\,mm. As expected, for small fibres the peak transmittance is close to 1 and the effective finesse is equal to the reflective finesse. For larger fibres, the transmission at the peak decreases and the width increases, which is equivalent to a reduction of the effective finesse. Simultaneously, the peak wavelength shifts towards a shorter wavelength, which reflects the fact  that the number of rays having a incidence angle $\theta>0$ has increased. Interestingly, the peaks remain symmetric, even for very large fibres, which is due to the fact that we have assumed uniform illumination and a circular fibre. This assumption is justified by the fact that the primary source we are using has a larger \'etendue than the fibre. The fibre is over-filled by the source and uniformly illuminated. In the real system we are using an octagonal fibre that has even better scrambling properties and produces increased illumination uniformity. \citep{Chazelas2010}.\\

\begin{figure}[htbp]
\centering
\includegraphics[width=9.5cm,height=7cm]{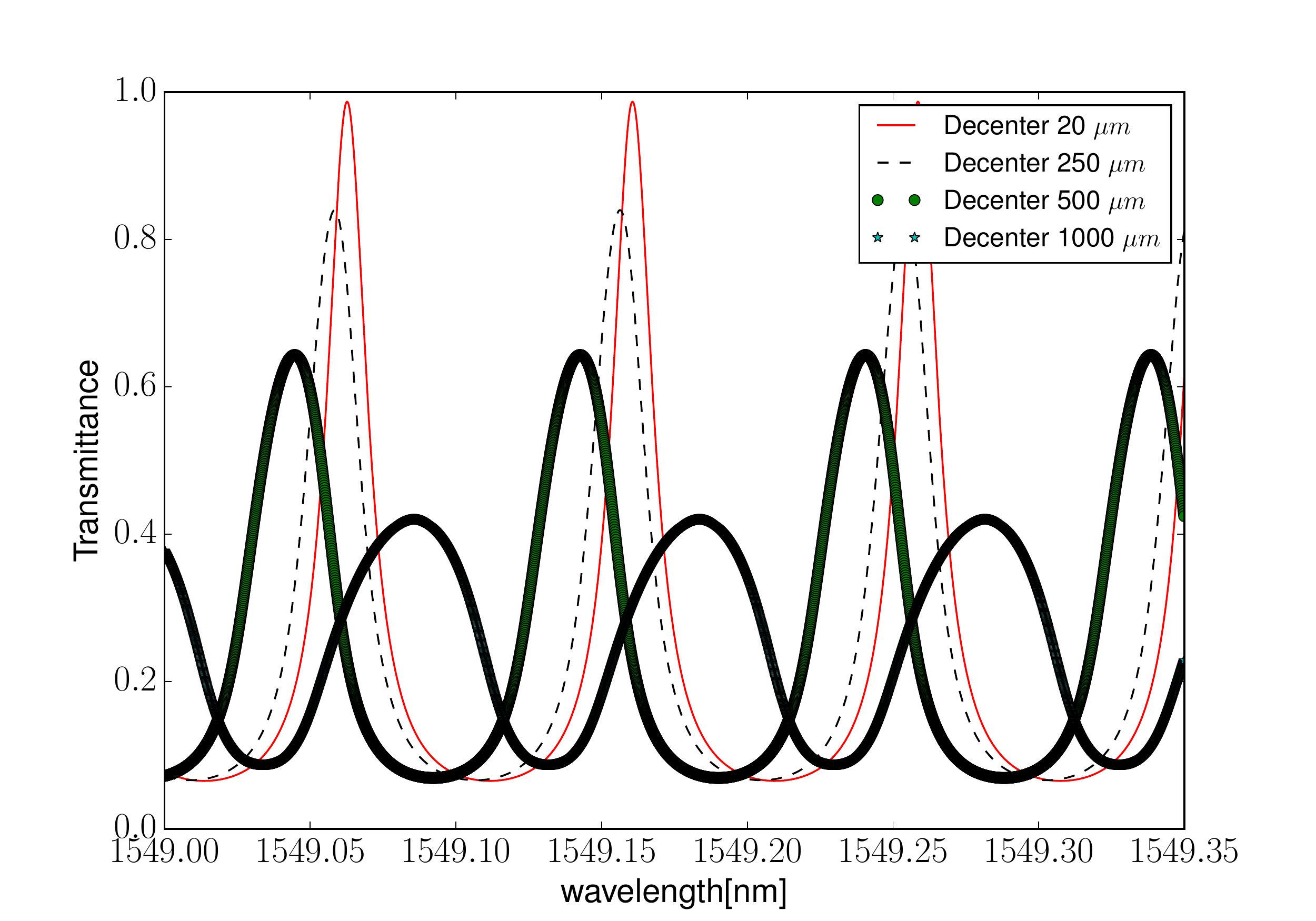}
\caption{Simulated transmittance of a Fabry-P\'erot for fibre de-centering value of 20, 250, 500, and 1000\,$\mu m$ assuming a fixed fibre diameter of $D=200$\,$\mu m$ and a focal length of the collimator $f=100$\,mm }
\label{fig:5}
\end{figure}

Figure \ref{fig:5} shows the results of the simulation as a function of the fibre de-centring, in which we fixed the fibre diameter to the chosen value of $D=200$\,$\mu$m. In this case we also observe a reduction of finesse and a blue-shift of the peaks with increasing de-centring. It is interesting to note, however, that when combining a finite fibre size with de-centring, the peaks become asymmetric.\\ 

\begin{figure}[htbp]
\centering
\includegraphics[width=10cm,height=5cm]{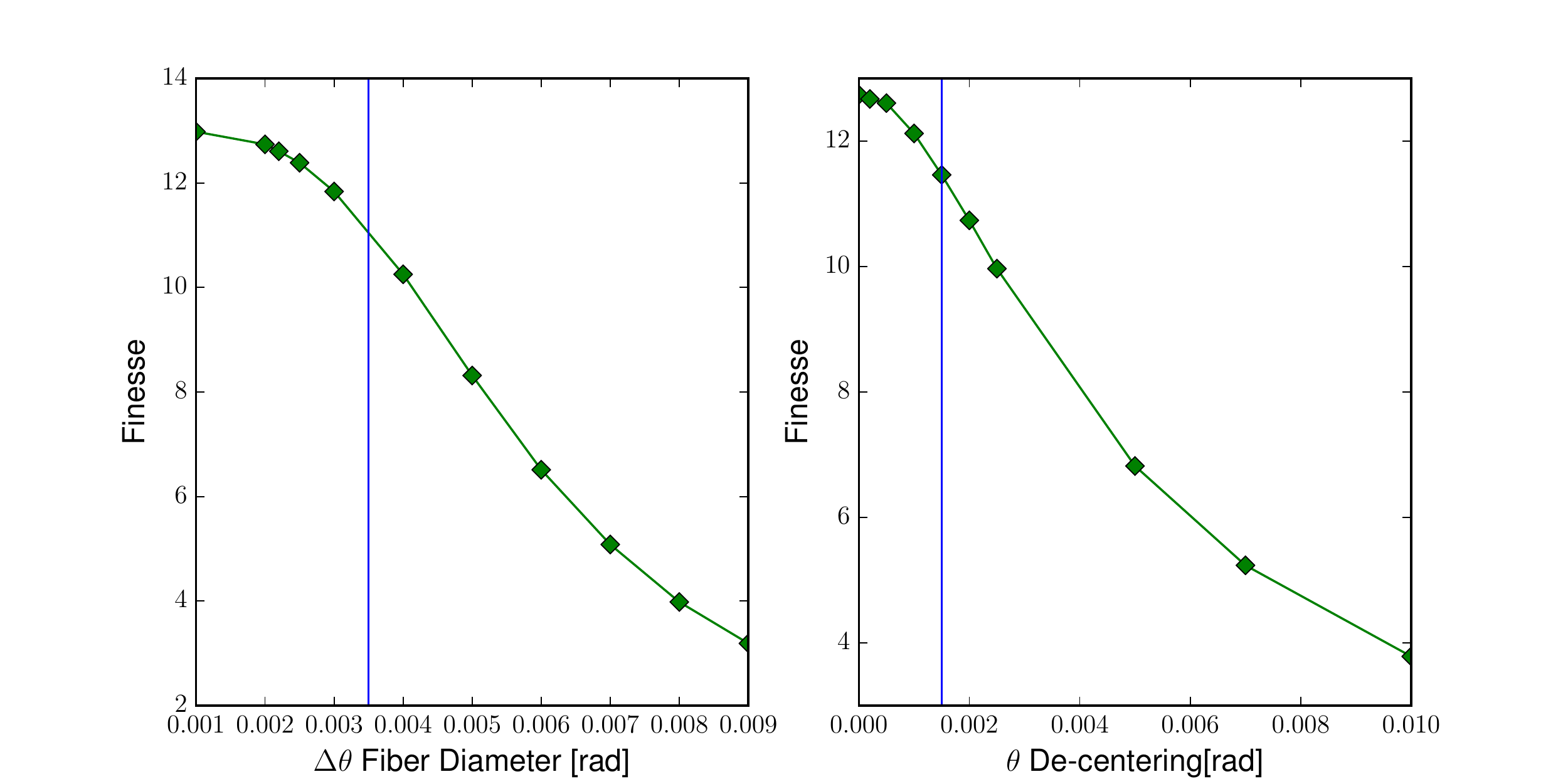}
\caption{Finesse for a centred fibre as a function of fibre diameter (left) and as a function of fibre de-centring (left) for a fixed fibre diameter $D=200$\,$\mu$m. The blue line marks the area we are interested in namely a small fibre diameter and an effective finesse close to the nominal value $F_R=13$.}
\label{fig:6}
 \end{figure}

The results for the finesse are summarized in Figure \ref{fig:6}. The plot shows the evolution of the finesse as a function of fibre diameter (left hand) and of the fibre de-centring (right hand). To be more general, we have expressed both, the fibre diameter and de-centring, in terms of angular values $\Delta\theta=D/f$ and $\theta=\Delta/f$, respectively. In the case of a small fibre diameter or small de-centring, the value for the effective finesse approaches the nominal value of $F_{R}=13$. It is important to note that the effective finesse remains larger than $F_{Eff}=11$ for a fibre diameter up to $\Delta\theta=0.0035$ ($D=350$\,$\mu$m), which perfectly justifies our choice for a $200$\,$\mu$m. However, the plot as a function of de-centring also tells us that, if we want to ensure the required effective finesse, the fibre must be centered to better that $\theta=0.0015$ ($\Delta=150$\,$\mu$m) for the optical axis of the system.

As mentioned before, the peak wavelength shifts towards the blue when increasing the fibre diameter or de-centring the fibre. When the Fabry-P\'erot is used for absolute wavelength calibration, the position of the peaks must be known. Even more critical is that, when used as a RV-reference (to measure the instrumental drift between two calibration sequences), we have to make sure that the transmission peaks remain stable in wavelength. Their dependence on fibre diameter and centring will set the  requirements on the focus and centring stability of our optical system. For this reason, we have expressed the wavelength shift in terms of 'radial-velocity' in the unit of [m\,$s^{-1}$]. The conversion from wavelength to speed is obtained by the relation:

\begin{equation}
\label{equa:DV}
\frac{\Delta v}{c}  = \frac{\Delta \lambda}{\lambda}
\end{equation}

\begin{figure}[htbp]
\centering
\includegraphics[width=9.5cm,height=4.5cm]{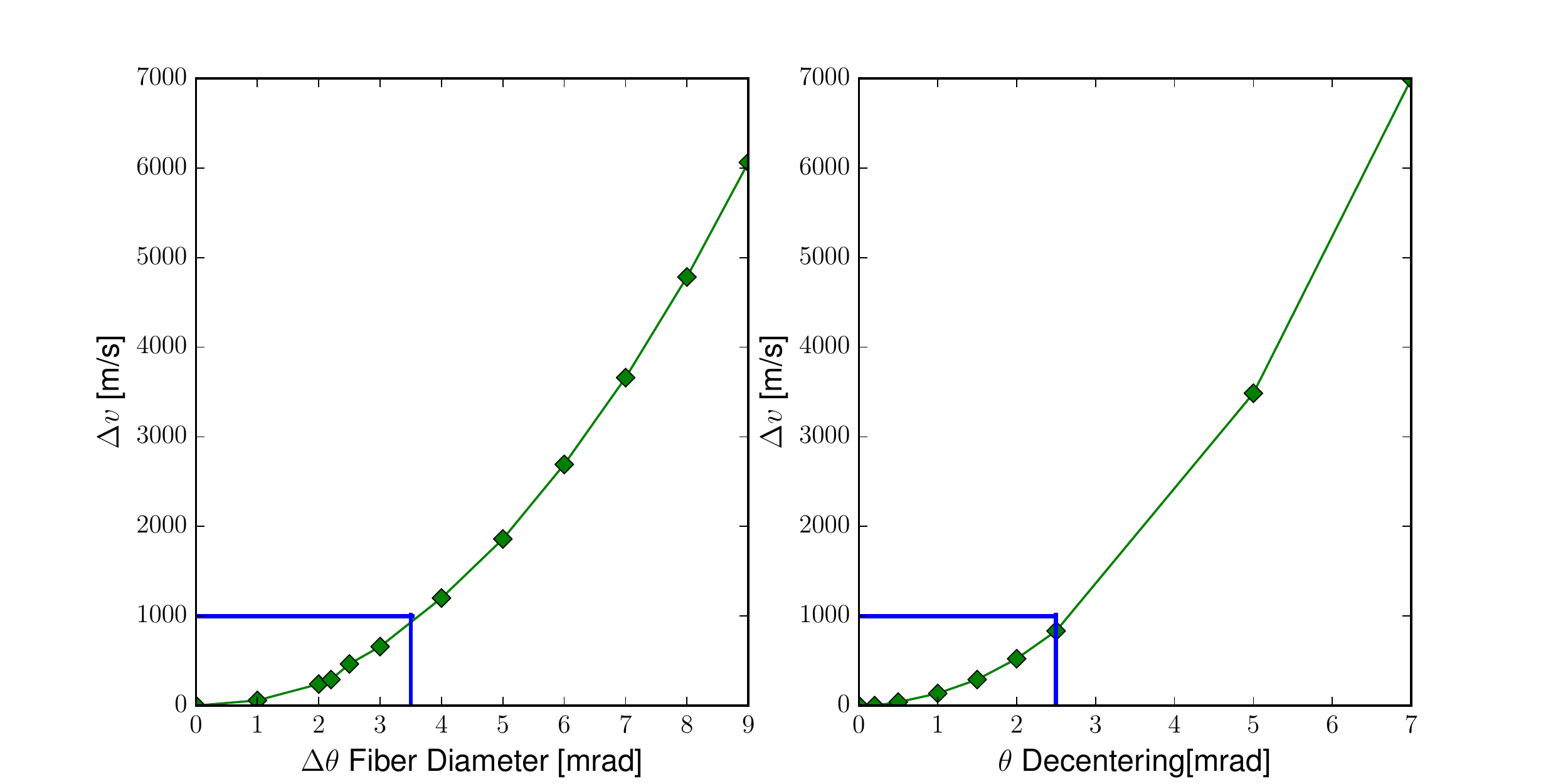}
\caption{Change in radial velocity for a centred fibre as a function of fibre diameter (left) and as a function of fibre de-centring (left) for a fixed fibre diameter $D=200$\,$\mu$m. The blue line marks the limit area we are interested in. The area is small because small changes introduces shift of several hundreds  or more of $m\,s^{-1}$ in both cases.}
\label{fig:7}
 \end{figure}

\begin{figure}[htbp]
\centering
 \includegraphics[width=9.6cm,height=4.5cm]{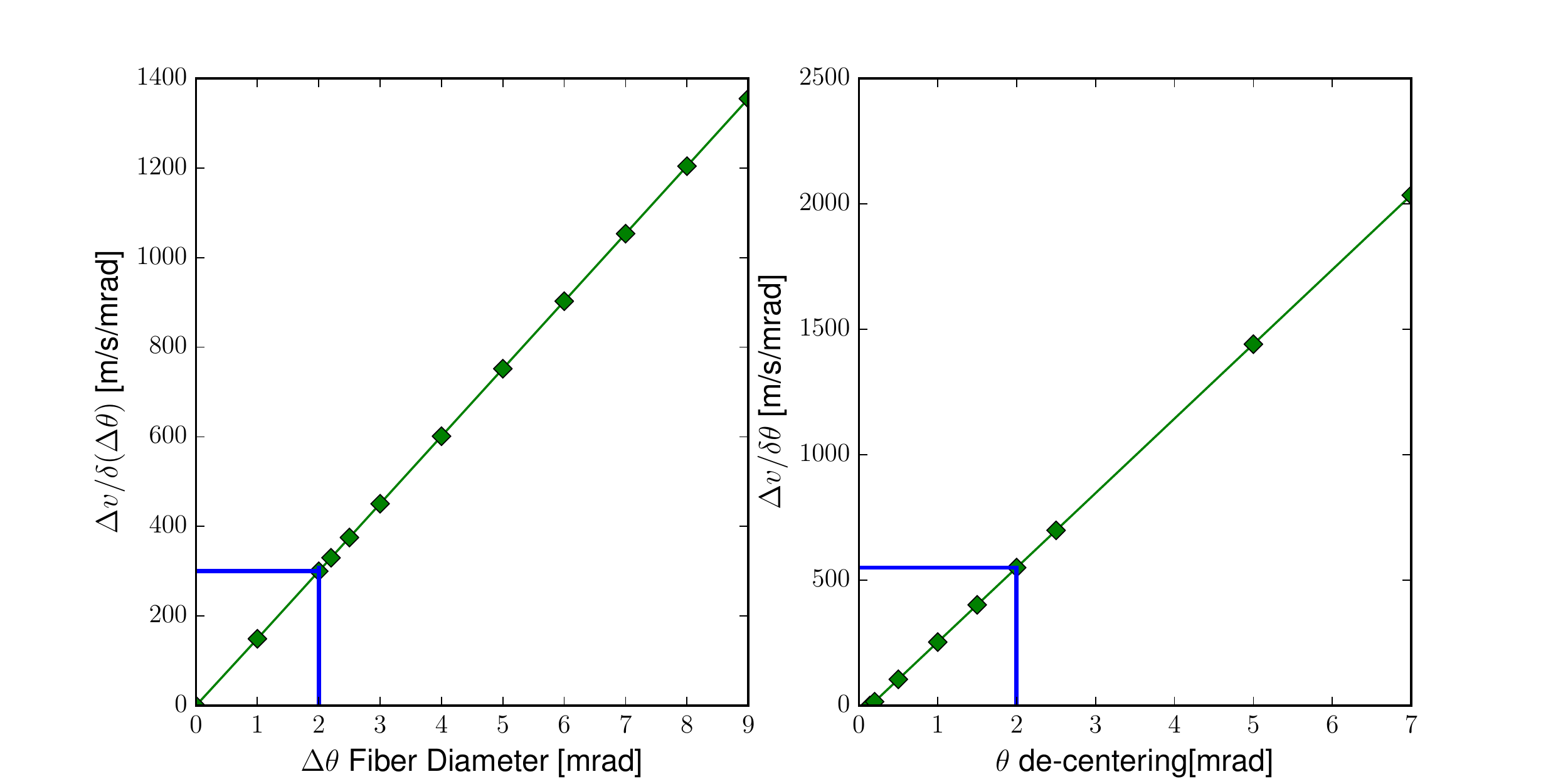}
\caption{Radial velocity sensitivity to fibre-size change for a centred fibre (left) and to de-centring for a fixed fibre size of $D=200$\,$\mu$m (right). The blue line marks the limit area we are interested in. For 2 mrad fibre size, the RV will change by 300 m\,s$^{-1}$ (3m\,s$^{-1}$ micron$^{-1}$) for a centred fibre. In the case of de-centring, it changes by 600 m\,s$^{-1}$ (6\,m\,s$^{-1}$  micron$^{-1}$).}
\label{fig:8}
\end{figure}

Figure \ref{fig:7} presents the shift of the transmission peak as a function of similarly to Figure \ref{fig:6} fibre diameter $\Delta\theta=D/f$ and de-centring $\theta=\Delta/f$. It can be seen that even a small change introduces shifts of several hundreds or even thousands of m\,s$^{-1}$. However, there is no reason besides thermo-mechanical instabilities for the fibre diameter or the de-centring to change with time. To better evaluate the calibration sensitivity to these parameters, we chose to presents the derivative of these curves (Figure \ref{fig:8}). Since the curve has a parabolic shape (as can be understood from the fact that $cos\theta \sim \theta^2$ for small angles $\theta$), its derivative is nothing more than a slope. The left-hand plot simply shows us that, the larger the fibre, the more sensitive the system is to a de-focus, and, from the right-hand plot, the same conclusion can be drawn for the de-centring. Given the chosen fibre size (2\,mrad), we derive that the radial velocity will change by 300\,m\,s$^{-1}$ m\,rad$^{-1}$. In physical terms for our system, this is 3\,m\,s$^{-1}$ micron$^{-1}$ (relating to the change in fibre size). In terms of de-centring, the sensitivity will approach zero if the fibre is perfectly centred and will increase to 600\,m\,s$^{-1}$ mrad$^{-1}$ (6\,m\,s$^{-1}$ micron$^{-1}$) of additional de-centring for a fibre de-centred by 2\,mrad (200\,$\mu$m).\\

From this analysis we conclude that small fibres and excellent alignment are mandatory not only for optimum finesse and flux, but also to reach the highest spectral stability. Below, we use the derived sensitivities to specify the thermo-mechanical requirements of the SPIRou RV-reference module.

\subsection{Opto-mechanical requirements derived from stability requirements}

\subsubsection{Alignment precision}
The analysis above shows that the RV-sensitivity of the \'etalon to opto-mechanical mechanical misalignment increases with de-centring. A perfectly-aligned system would not be sensitive to motions of the fibre, for example due to thermal or gravity effects, but even for slight misalignment our system could become unstable at radial velocities beyond 30\,cm\,s$^{-1}$. Assuming that we can align the fibre by auto-collimation to better than one-twentieth of its diameter, that is to 10\,$\mu$m and using the results presented in Figure \ref{fig:8}, we obtain a sensitivity of 30\,m\,s$^{-1}$ mrad$^{-1}$ (0.3\,m\,s$^{-1}$ micron$^{-1}$). In other words, the fibre position must be stable to within one micron of the focal plane of the collimator and the angular alignment of the \'etalon with the collimators must not change by more than 0.01\,mrad.

\subsubsection{Pressure stability}
\label{pressstab}
For an air-spaced \'etalon like SPIRou's, any pressure change will induce a change in the optical gap and thus in the wavelength of the transmitted peaks. For this reason we made the choice to operate the RV-reference module in a vacuum (see Figure \ref{fig:9}). To establish whether the system must be pumped continuously or not, we have to compute the drift of the system arising from pressure changes that have occurred since the last wavelength calibration of the spectrograph, which taking place at the start of the night. In other words, pressure changes in the Fabry-P\'erot system must not induced line-position changes larger than the required RV-precision. Given the requirement for the RV-precision of 30\,cm\,s$^{-1}$ during an observing night (typically 12 hours) and using Eq.\,\ref{equa:FP} and Eq.\,\ref{equa:DV}, we computed that the refractive index $\Delta n/n$ must not change by more than $10^{-9}$ in relative terms. Using the approximation for the air index $n_{air} -1 = 0.00027 \cdot p/p_{atm}$, where $p_{atm}$ is the average ambient pressure at sea level, we computed that the pressure $p$ inside the vacuum tank must not change by more than $3.7 \cdot 10^{-3}$\,mbar in 12 hours or $7.4 \cdot 10^{-3}$\,mbar day$^{-1}$.\\

\subsubsection{Operating pressure}
\label{operpress}
An aspect which is frequently forgotten, is that pressure does not only depend on air leaks but on the volume of the air contained in the vacuum tank. External ambient pressure changes due to meteorological effects affect the volume of our vacuum tank due to the variation in  pressure differences between outside and inside. Consequently, the index of the residual air inside the vacuum tank will vary and degrade the calibration system. The sensitivity to atmospheric pressure changes depends on the air pressure (or density) inside the vacuum tank. We must therefore understand at which (maximum) pressure we must operate the RV-reference module to avoid effects larger than 30\,cm\,s$^{-1}$.\\

Bearing in mind Boyle's law $p \cdot V= const $, defining $V_{0}$ as the initial volume, and defining $p_0$ the initial pressure inside the vacuum tank, the refractive index $n$ of the residual air can be written as:
\begin{equation}
n-1=\frac{p}{p_{Atm}}\cdot (n_{Atm}-1) = \frac{V_0}{V}\cdot \frac{p_0}{p_{Atm}}\cdot (n_{Atm}-1)
\end{equation}

where $(n_{Atm}-1)\approx 2.7 \cdot 10^{-4}$ corresponds to the refractive index of standard air at atmospheric pressure $p_{Atm}$, and $p$ and $V$ are the actual pressure and volume inside the vacuum tank.\\

The derivative of the refractive index $n$ with respect to the change of the external pressure $p_{out}$ can then be written as:
\begin{equation}
\frac{dn}{dp_{out}}=\frac{dn}{dV}\cdot\frac{dV}{dp_{out}}= \beta \cdot \frac{p_{0}}{p_{Atm}}\cdot (n_{Atm}-1)
\end{equation}

where $\beta:=-\frac{dV}{V \cdot dp_{out}}$ is the relative volume change of the vacuum tank as a function of changes in external pressure (at constant internal pressure).\\

By means of a finite element analysis of the vacuum tank (the following sections detail the design description), we computed $\beta \sim 10^{-5}$\,mbar$^{-1}$. This value allows us to determine the maximum operational pressure to avoid change in the radial velocity of the RV-reference module larger than the specified 30\,cm\,s$^{-1}$. If we consider that ambient pressure variations hardly exceed 10\,mbar night$^{-1}$ at 2500\,m altitude (e.g. at the La Silla and Paranal Observatories), we derive an upper limit for the operation pressure of $p_{0}<30$\,mbar. In practical terms, and given the maximum allowed pressure-rise rate of d$p/$d$t = 7.4 \cdot 10^{-3}$\,mbar day$^{-1}$ determined in the previous section, it is sufficient to evacuate the RV-reference module once every 10\,years in the worst case.\\

\subsubsection{Temperature stability}\label{opertemp}
We considered two effects induced by temperature variations. The first is the thermal stability of the optical gap that is essentially given by CTE of the spacers. The spacer material is from Corning ULE and thus the CTE is guaranteed to be lower than $2 \cdot 10^{-8}$ Kelvin$^{-1}$. This translates into a radial-velocity sensitivity of about 6.7\,m\,s$^{-1}$\,K$^{-1}$. Again applying the radial-velocity stability requirement, we derive a temperature stability requirement for the RV-reference module of better than 0.05\,K.\\

The second temperature effect may produce focal-length changes of the collimator that would result in a 'de-focus' of the entrance fibre and an apparent increase of fibre diameter $D$. As we will describe in Sec.\,\ref{par:design}, the Fabry-P\'erot assembly of our RV-reference module is made of aluminium (also Figure \ref{fig:9}). Considering the focal length of $f=100$\,mm and the coefficient of linear thermal expansion for aluminium as $CTE_{Al}=2.3 \cdot 10^{-5}$\,K$^{-1}$, we computed an absolute focal length change of d$f/$d$T = 2.3$\,$\mu$m\,K$^{-1}$. Because the F-number of the fibre exit (or collimator) is F/2.5, we obtained an increase of apparent diameter d$D/$d$T \sim 1$\,$\mu$m\,K$^{-1}$. From the plot in Figure \ref{fig:8} we can then derive that, for a fibre of $200\mu$m diameter, the radial velocity would change by $\Delta v \sim 3 m/s$\,K$^{-1}$. In order to comply with the radial-velocity requirements, we have to keep the temperature of the system stable to 0.1\,K.\\ 

From the analysis above, we conclude that it is sufficient to control the temperature of the RV-reference module to a level of 0.05\,K even when using aluminium for its structure. This kind of temperature stability is easily achieved with commercial components. Nevertheless, we made a particular effort to design the opto-mechanical system in the most symmetric and stable way, and to thermally isolate the critical parts inside the vacuum from the vacuum tank. Therefore operating in a vacuum will further improve the thermal stability.

\subsection{Design of the SPIRou radial-velocity-reference module}\label{par:design}
\begin{figure}[htbp]
\centering
\includegraphics[width=8cm, angle=0]{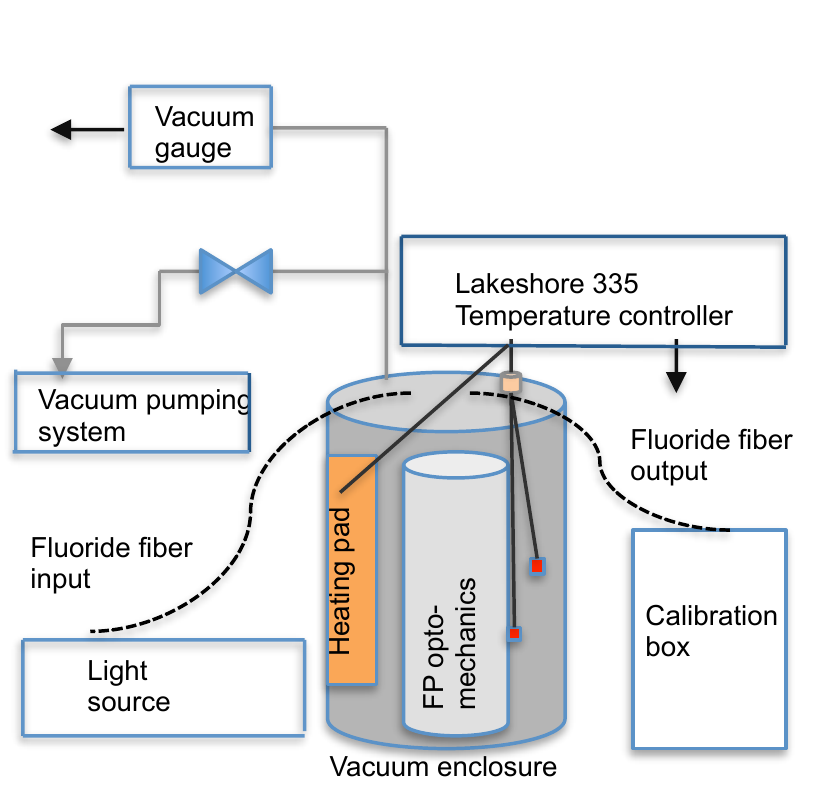}
\caption{Functional diagram of the RV-reference unit}
\label{fig:9}
\end{figure}

The RV-reference module designed for SPIRou (Figure \ref{fig:9}) is derived from a concept that was successfully developed for HARPS and the visible wavelength range \citep{Wildi2010}. The design was later made more modular and adapted for the infrared wavelength range by replacing lenses with mirrors. This was successfully applied firstly to the HARPS-N and CORALIE modules and was later essentially copied for SPIRou.\\

The system conceptually consists of a Fabry-Perot \'etalon filter illuminated in white light to produce a set of equally-spaced (in frequency) emission lines of uniform intensity across a wide wavelength range. For the illumination we use a laser-driven light source (LDLS by Energetix) that essentially produces a black-body emission of 10'000\,K. Although at visible wavelengths the surface brightness of this sort of source is considerably higher than a standard halogen source of about 3'000\,K, the brightness difference at wavelengths above 1\,$\mu$m is much lower and almost negligible. For our system we nevertheless decided to use an LDLS with a 200\,$\mu$m fibre output that perfectly matches the \'etendue of the \'etalon. The LDLS light is fed to the RV-reference module through an octagonal fluoride fibre of 300\,$\mu$m core diameter and $2\,$m in length, whereas the RV-reference module output fibre, which collects the light from the module and brings it to the spectrograph, is of 600\,$\mu$m core diameter and $4\,$m length.

\begin{figure}[htbp]
\includegraphics[scale=0.35]{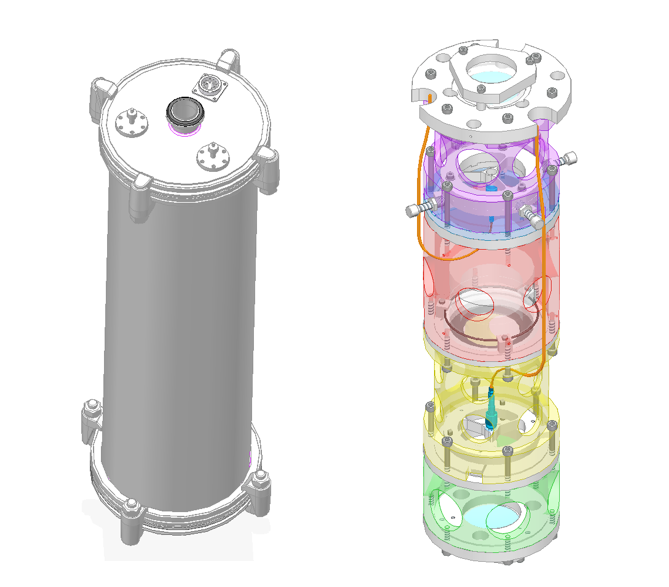}	\caption{View of the vacuum tank (left) and the assembly in which the Fabry-P\'erot, fibres and mirrors are installed on the mechanical support structure (right).}
\label{fig:10}
\end{figure}

A vacuum tank contains the fibre-fed Fabry-P\'erot opto-mechanical system (Figure \ref{fig:10}). The pressure is initially pumped to $10^{-4}$\,mbar or less, but eventually the valve is closed. More important than the actual pressure value is the pressure rise rate which must comply with the previously mentioned requirements. The temperature is controlled using heating-resistor foils glued around the vacuum tank. Two precise silicon-diode temperature sensors (one on the vacuum tank for the control and the other close to the \'etalon for monitoring) and a Lakeshore 331 device close the temperature-control loop to better than 5\,mK \emph{rms}. The whole system is isolated and packed inside a box to avoid heat exchange with the laboratory environment.\\

The optical system is placed in the vacuum vessel and fixed to the top flange in only 3 points, using thermally insulating material to minimize thermal conduction from outside. The collimating and focusing optics comprise identical protected silver-coated parabolas with a focal length of 100\,mm. Parabolic mirrors were selected because they result in low aberrations and guarantee that the line profile stays symmetric. 

The light enters the vacuum tank through an octagonal silica fibre purchased from CERAMOPTEC (OPTRAN WF). Its diameter is 200\,$\mu m$ in accordance to the derived finesse requirements. The octagonal shape was chosen to ensure a very uniform and thus stable illumination of the \'etalon \citep{Chazelas2012}. This is important to minimize possible illumination effects at the source level, such as PSF changes and variations of the illumination spot position or its angular distribution.\\

The connectorized fibre is mounted on a spider and illuminates, on its axis, a parabolic mirror of 50\,mm clear aperture. Given the 100\,mm focal length of the mirror, the F/2.5 beam of the fibre produces a collimated beam of 40\,mm diameter. The back-reflected beam is partially obstructed by the fibre and its spider, but we preferred to accept some light losses for the sake of symmetry. The beam  crosses the \'etalon before being re-focused by a second on-axis parabola on the output fibre, that is mounted in exactly the same way as the collimation part.\\

The output fibre inside the vacuum tank is circular and its diameter is 600\,$\mu m$. A larger fibre was chosen in order make alignment easier, collect all the light, and eventually increase the stability of the system. Being located after the \'etalon, the diameter of this fibre has no influence on the resulting finesse.\\

\begin{figure}[htbp]
\centering
\includegraphics[width=6cm,height=3.5cm]{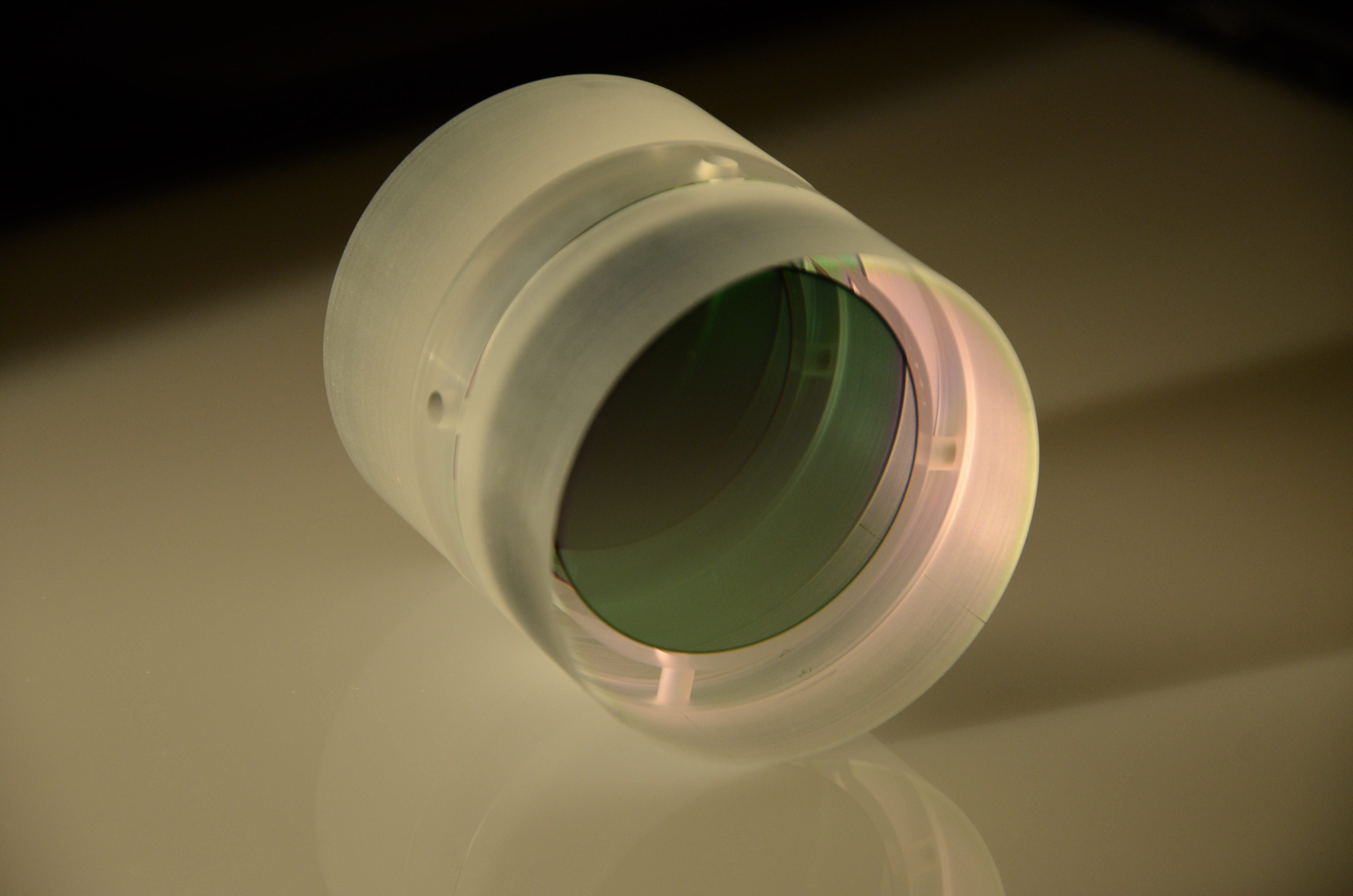}
\caption{Picture of the SPIRou Fabry-P\'erot etalon.} 
\label{fig:11}
\end{figure}

\begin{figure}[htbp]
\centering
\includegraphics[width=9.5cm,height=4cm]{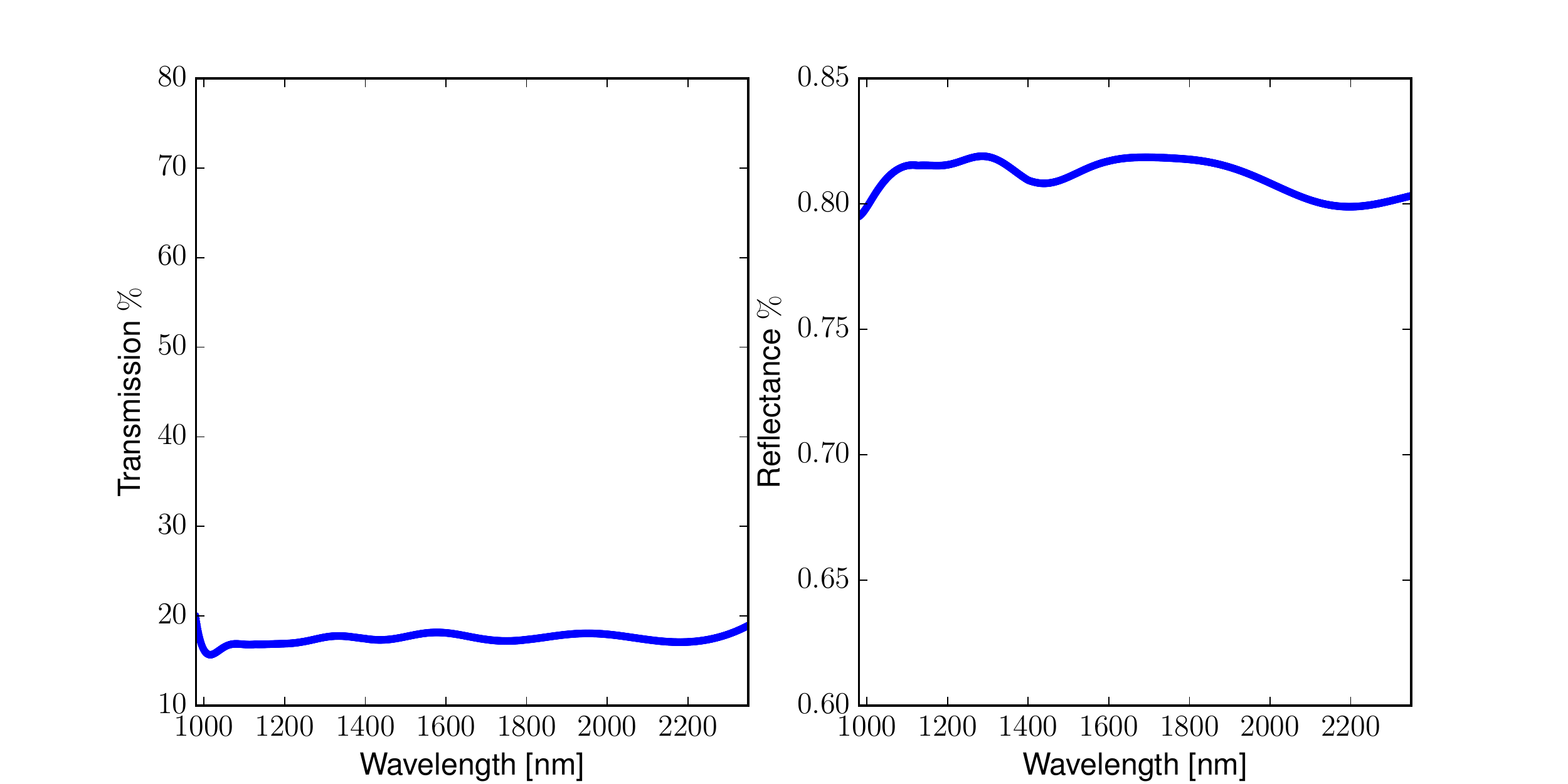}
\caption{Transmission (left) and reflectance (right) of an etalon mirror}
\label{fig:12}
\end{figure}

The Fabry-P\'erot etalon used in SPIRou is shown in Figure \ref{fig:11}. Its physical diameter is of 60\,mm while the clear aperture is 50\,mm. The two mirrors are made of Suprasil. They were polished and coated by the company Thin Films Physics, Switzerland. The surface flatness is of the order of $\lambda/100$. The coating reflectivity of $R=80\%+2\%$ over the $980-2350\, nm$ range was designed to match the reflectivity-finesse requirements (Figure \ref{fig:12}). Both mirrors were anti-reflection (AR) coated on the external face. Nevertheless, a wedge was introduced between the two faces of each mirror to avoid possible parasitic interference with the external faces. The two mirrors are separated by a set of three high-precision Corning ULE (Ultra Low Expansion material - $CTE < 2x10^{-8}$\,K)) spacers produced by sls-optics, UK. The \'etalon was finally assembled using dry optical contacting by ICOS, UK.


\section{Laboratory test}\label{ref:four}
\subsection{Test goals}

We characterized the RV-reference module in the laboratory with respect to its basic parameters. It turned out to be quite a demanding task to measure the high-resolution transmission function because of the high spectral resolution of the \'etalon itself. To sufficiently sample the line profile we would have needed a spectrograph or spectrometer of resolving power $R>500'000$ that only a Fourier-Transform-Spectroscopy (FTS) can deliver. We decided to restrict ourselves to a tunable laser wavelength for this measurement. Once spectral performances are demonstrated at one wavelength, and knowing the \'etalon gap and the broad-band reflectivity, the performances can be determined also at other wavelength. Nevertheless, we have separately determined the broad-band transmittance and the flux delivered by the system. Also, we have measured the pressure and temperature stability to verify that the system complies with the technical requirements. 

\subsection{Measurements $\&$ results}
\subsubsection{Finesse}
\begin{figure}[htbp]
\centering
\includegraphics[width=8cm,height=5.8cm]{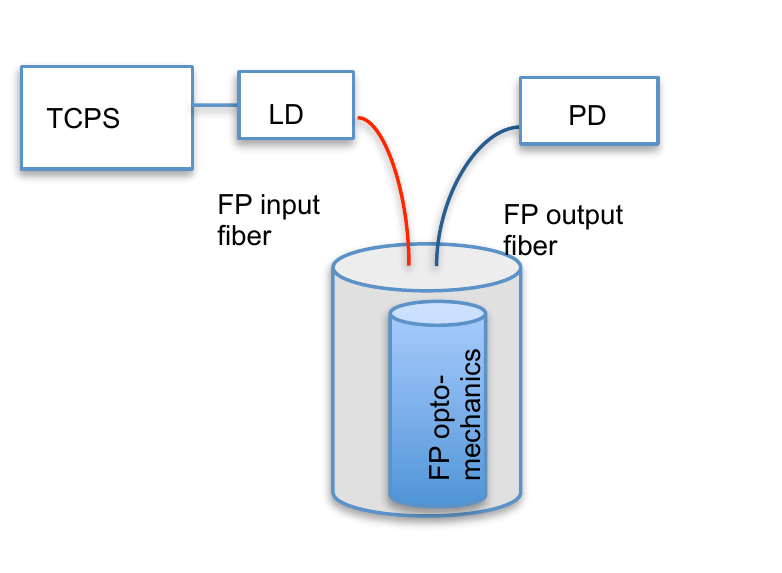}
\caption{Principle of our test set-up. TCPS: temperature controller and laser diode power Supply; LD: laser diode; FP: Fabry-P\'erot opto-mechanics, PD: photodiode}
\label{fig:13}
\end{figure}

Figure \ref{fig:13} shows the set-up used for characterizing our \'etalon. We used a temperature-controlled DFB laser diode at 1540\,nm as a tunable laser. The centre wavelength has a temperature coefficient of 0.095\,nm\,K$^{-1}$. We used this characteristic to scan the transmittance of the Fabry-P\'erot as a function of wavelength. The source's spectral width ($FWHM =2$\,MHz) is narrow enough to probe the line profile of our \'etalon ($FWHM =1$\,GHz). The flux was measured at the output of the system by an indium-gallium-arsenide (InGaAs) photodiode.\\

\begin{figure}[htbp]
\centering
\includegraphics[width=9.5cm, height=7cm]{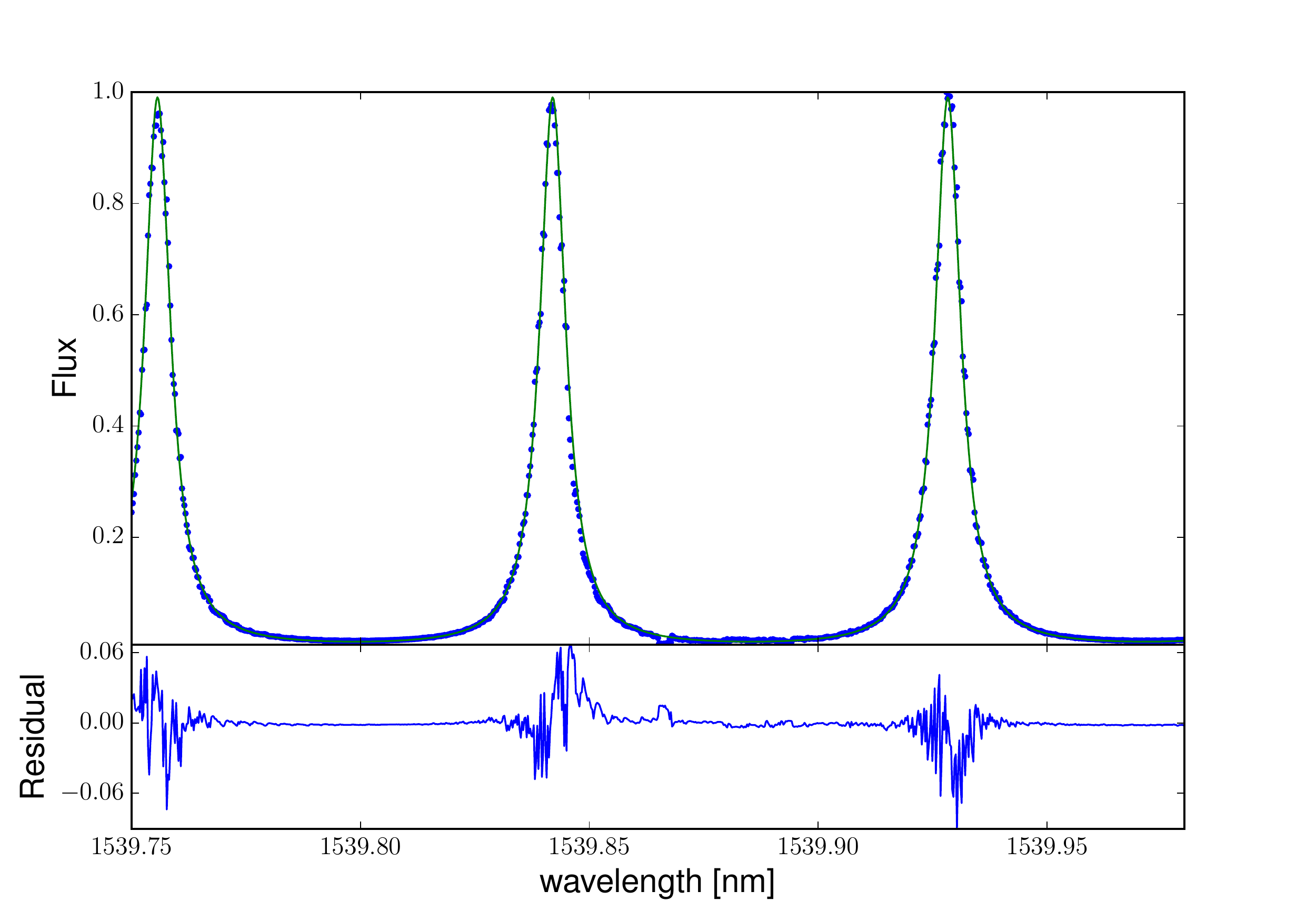}
\caption{Measured Fabry-P\'erot transmittance as a function of wavelength (blue dots) and fitted theoretical curve (green line). On the bottom the residuals to the fit are shown.}
\label{fig:14}
\end{figure}

Figure \ref{fig:14} shows the measured transmittance curve $T(\lambda)$ of the RV-reference module. The finesse can be estimated by dividing the free-spectral range (distance between two neighbouring peaks) by the $FWHM$ of a single line. Alternatively, we fitted the obtained curve by a theoretical transmission function (green curve in Figure \ref{fig:14}) leaving the $FSR$ and the effective finesse $F_E$ as free parameters. The so obtained values are $F_E\sim 12.76$ and $FSR = \Delta\lambda=\lambda^{2}/2nD \sim 0.09$\,nm, both perfectly in line with the design values. The finesse is actually slightly higher than expected because of an actual coating reflectivity of about 81-82\% instead of the specified 80\%. Finally, we would like to note the exact match of the fitted curve with the measured data, which demonstrates that we have all the effects which could influence the line symmetry, such as the divergence finesse, well under control.\\

\subsubsection{Transmittance and flux}

\begin{figure}[htbp]
\centering
\includegraphics[width=9.5cm, height=3.9cm]{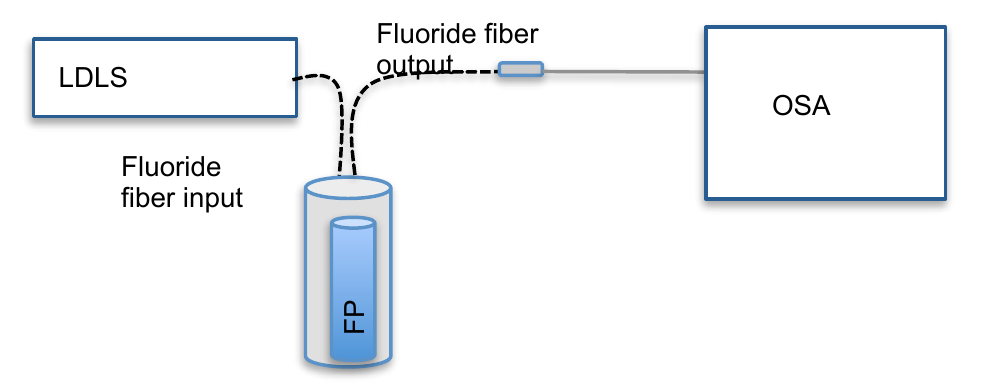}
\caption{Radial velocity reference module test set-up for spectral flux and
transmission measurements from 900\,nm to 1700\,nm. The setup includes an LDLS lamp, the Fabry-P\'erot system, a test fibre of $600\mu$m, and the optical spectral analyzer (OSA).}
\label{fig:15}
\end{figure}

\begin{figure}[htbp]
\centering
\includegraphics[width=9.5cm, height=3.9cm]{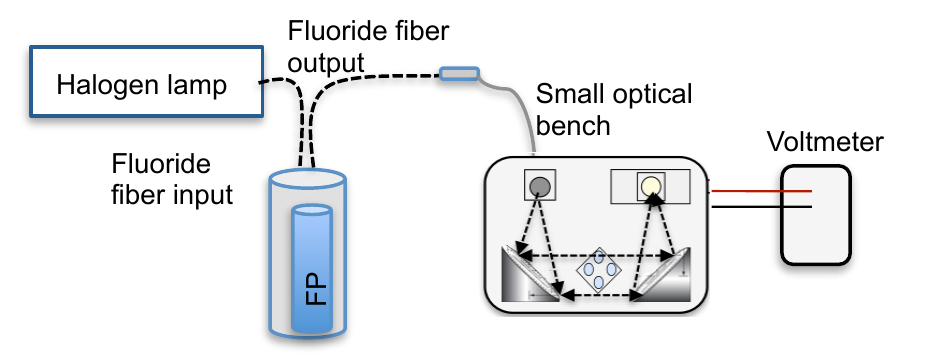}
\caption{Radial velocity reference module test set-up for spectral power and
transmission measurements from 1700 nm to 2460 nm. A small optical bench is included to analyze the light, comprising a collimator, filter wheel, focusing mirror, NIR photodiode, and voltmeter.}
\label{fig:16}
\end{figure}

We measured the spectral power and transmittance of the system in the $950-1700\,nm$ range directly with a laboratory Optical Spectrum Analyzer (OSA). Figure \ref{fig:15} shows the corresponding set-up. Figure \ref{fig:16} shows the set-up for the $1600-2400\,nm$ spectral range, for which we employed a small optical bench with selectable band-pass filters to measure the optical power and transmittance in four discrete bands by means of a NIR photodiode.\\

\begin{figure}[htbp]
	\centering
\includegraphics[width=10cm, height=7cm]{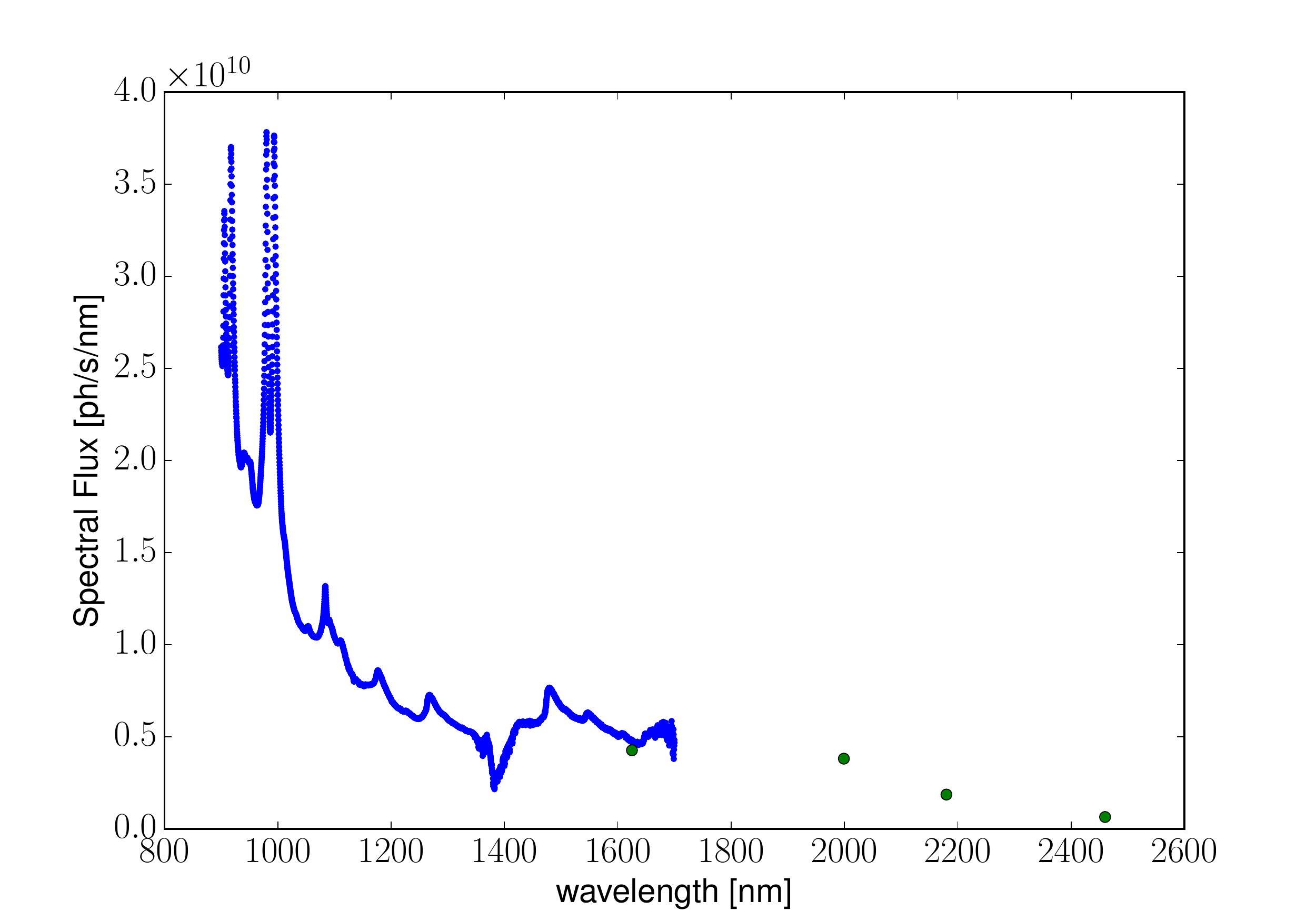}
\caption{Spectral flux of the RV-reference module when used with the LDLS. Beyond 1700 nm, only three discrete measurement points are available.}
\label{fig:17}
\end{figure}

\begin{figure}[htbp]
\centering
\includegraphics[width=10cm, height=7cm]{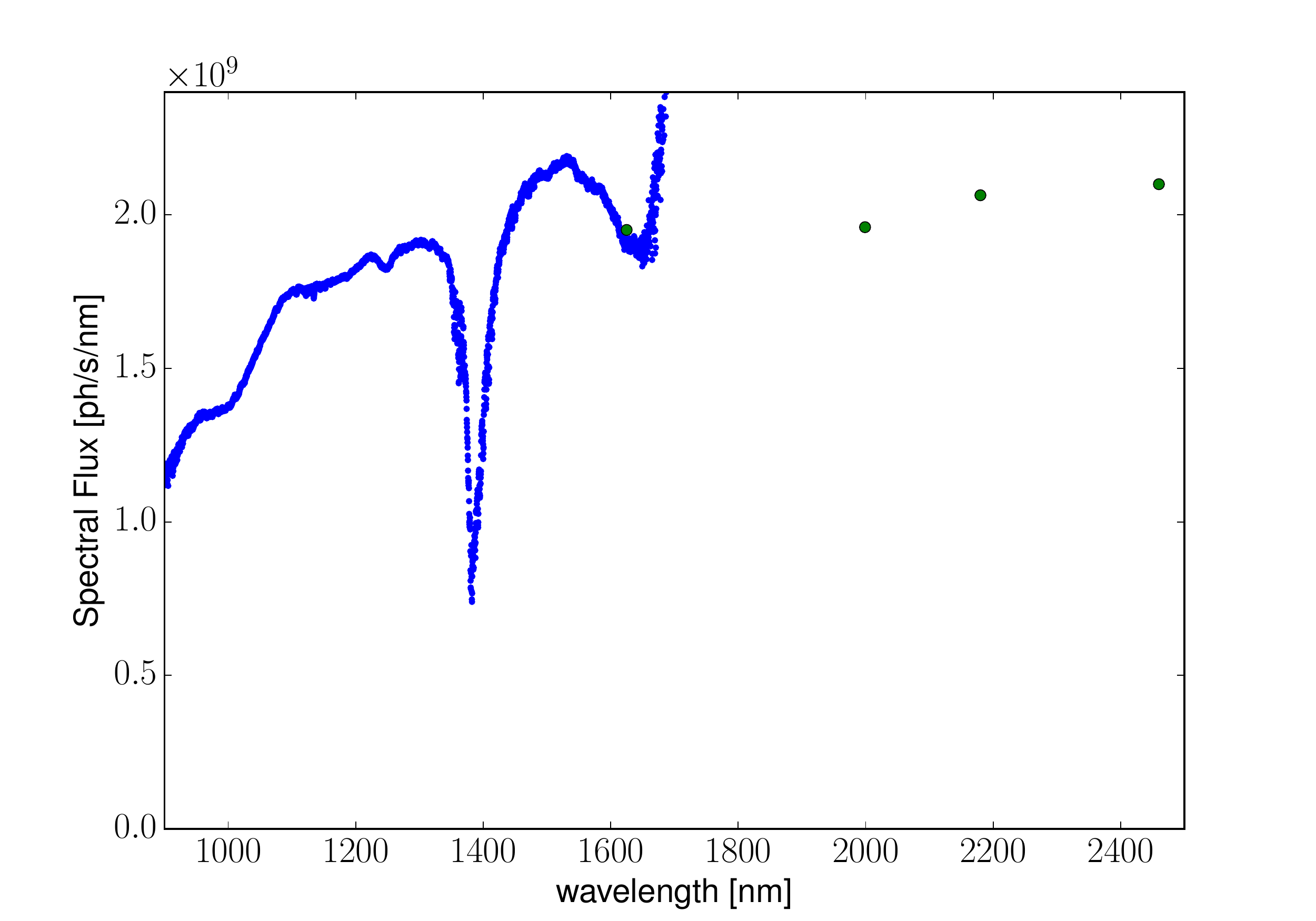}
\caption{Spectral flux of the RV-reference module when used with the halogen lamp }
\label{fig:18}
\end{figure}

The power was estimated using the internal calibrations and conversion factors of the OSA and the photodiode. The measurements should therefore not be considered to be of great accuracy but to provide a coarse estimate of the spectral-power density produced by the RV-reference module as a whole. Figure \ref{fig:17} shows the spectral flux produced by the RV-reference module as a function of wavelength. We originally observed a discrepancy of about a factor of 4.5 at $1625\,nm$ where we have overlapping measurements on the two set-ups. We approximately measured the same factor when using an halogen lamp instead of the LDLS to light-feed the module (Figure \ref{fig:18}). This factor can be explained by the bad match of the fibre diameter used to feed the OSA ($60\,\mu$m) with the RV-reference module output fibre. Not being able to correct for the coupling losses to the required precision, we decided to scale the measurements obtained with the OSA to the more accurate measurements with the silicon diode. To obtain the correct spectral flux we therefore multiplied the flux given for $\lambda <1600\,nm$ on Figure \ref{fig:17} and Figure \ref{fig:18} by the measured factor of 4.5. We should note that even then we obtain a lower limit for the actual flux because we neglected transmission losses in the measuring system itself, namely the mirror, diffuser, and sensor coupling.\\

In interesting to note that the flux above 1000\,nm produced by feeding the module with the two lamps (Figure \ref{fig:17} and Figure \ref{fig:18}), the LDLS or the halogen lamp, is not significantly different despite the very different temperature of the lamps. This is expected by the black-body emission law.  For cost considerations, we therefore used a halogen lamp in the final RV-reference module layout.\\

The transmittance of the RV-reference module was computed the ratio of the spectral powers measured at the output of the test set-up with and without the RV-reference module. In the latter case we simply connected together the input and output fibre of the module. The corresponding measurements for flux and transmittance in the four bands are summarized in Table \ref{tab:nofp}. The last column indicates the transmission of the RV-reference module at the peak $T_{FP_{p}}$ of the Fabry-P\'erot line, and has been computed by multiplying the average (broadband) transmittance of the system by the finesse of the Fabry-P\'erot.\\

\begin{table}[h]
\caption{Measured spectral power with $P_{FP}$ and without $P_{REF}$ the RV-reference module, and the average $T_{FP_a}$ and peak throughput $T_{FP_p}$  of the RV-reference module for four different filters of central wavelength $\lambda_c$ and bandwidth $\Delta\lambda$.} 
\centering 
\begin{tabular}{l l l l l l}
\hline\hline 
\textbf{$\lambda_c$} & \textbf{$\Delta\lambda$} & \textbf{$P_{REF}$} & \textbf{$P_{FP}$} & $T_{FP_{a}}$ & $T_{FP_{p}}$\\
\textbf{[nm]} & \textbf{[nm]} & \textbf{[W/nm]}&\textbf{[W/nm]} & \textbf{[\%]} & \textbf{[\%]}\\
\hline 
$1625.7$ & $64.1$ & $2.36$E-6 & $9.32$E-10 & 0.039 & 0.48 \\ 
$1999.5$ & $15.3$ & $1.43$E-6 & $5.88$E-10 & 0.041 & 0.58 \\
$2180.0$ & $15.0$ & $1.07$E-6 & $2.77$E-10 & 0.026 & 0.32 \\ $2460.8$ & $61.0$ & $6.15$E-7 & $4.37$E-11 & 0.0071 & 0.083\\
\hline\hline 
\end{tabular}
\label{tab:nofp}
\end{table}



\subsubsection{Pressure and temperature stability}
\begin{figure}[htbp]
\centering
\includegraphics[width=10cm, height=7cm]{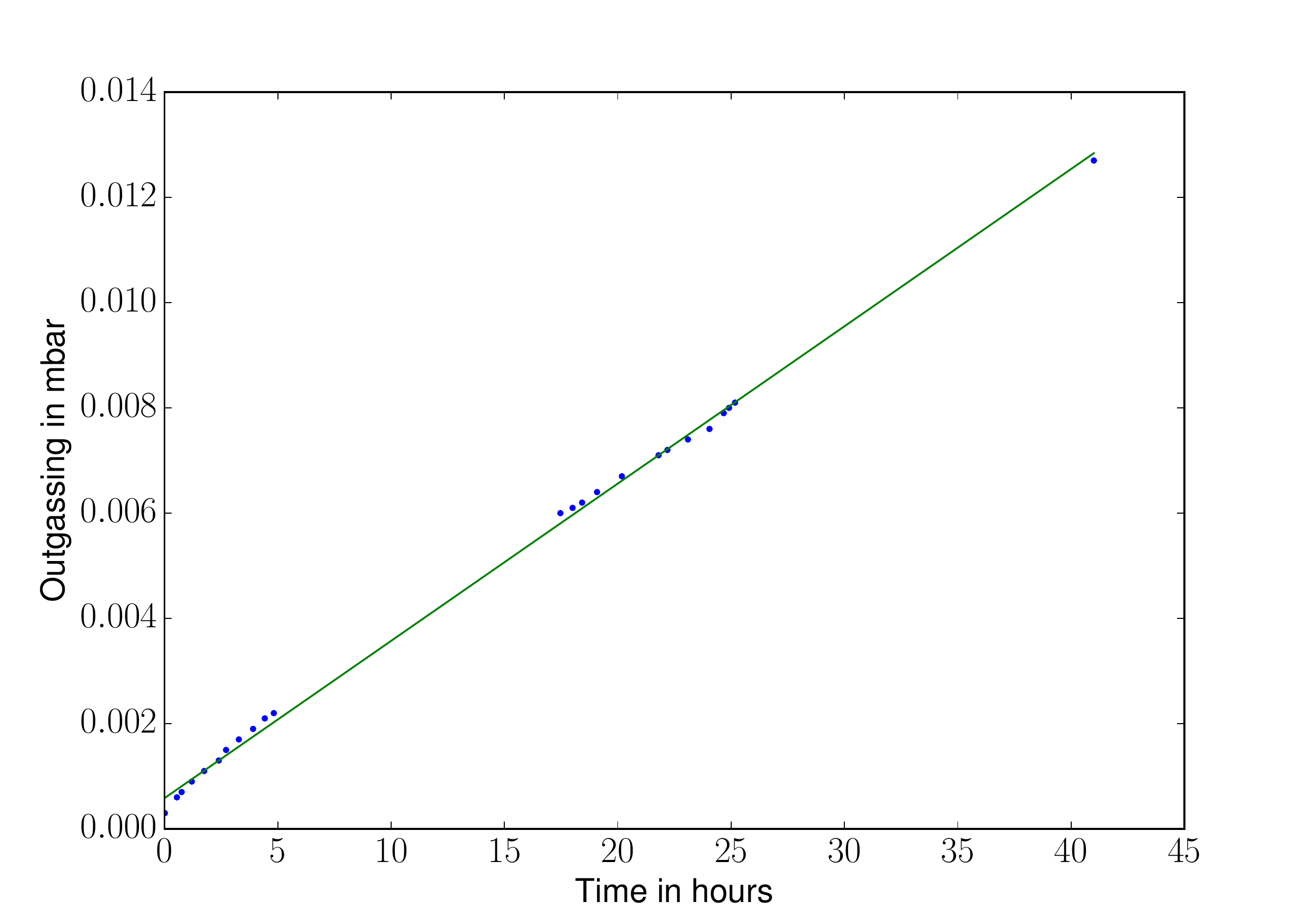}
\caption{Pressure (outgassing) as a function of time}
\label{fig:19}
\end{figure}

As explained in Sects. \ref{operpress} and \ref{opertemp} the RV-reference module must be operated in a vacuum well below 30\,mbar and the pressure raise rate must be lower than  $7.4 \cdot 10^{-3}$\,mbar. After evacuating the vacuum tank and doing a leak test to ensure the absence of vacuum leak, we performed a pressure test as shown in Figure \ref{fig:19}. The residual out-gassing rate is $7.2 \cdot 10^{-3}$\,mbar day$^{-1}$, just within the specification. Experience shows that out-gassing decreases as a function of time to a level much better than the required value.

\begin{figure}[htbp]
\centering
\includegraphics[width=5.5cm,height=5.5cm]{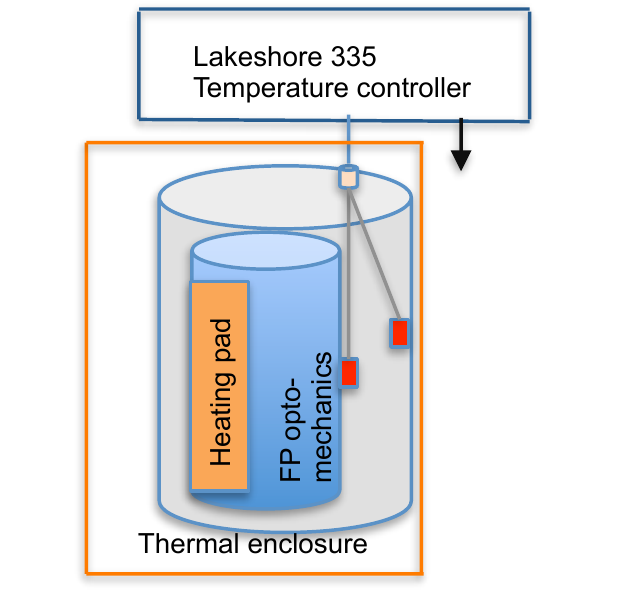}
\caption{Scheme of the temperature control}
\label{fig:20}
\end{figure}

We tested the temperature stability in a similar way. Figure \ref{fig:20} shows the diagram of the set-up for temperature monitoring.

\begin{figure}[htbp]
\centering
\includegraphics[width=9cm,height=6.5cm]{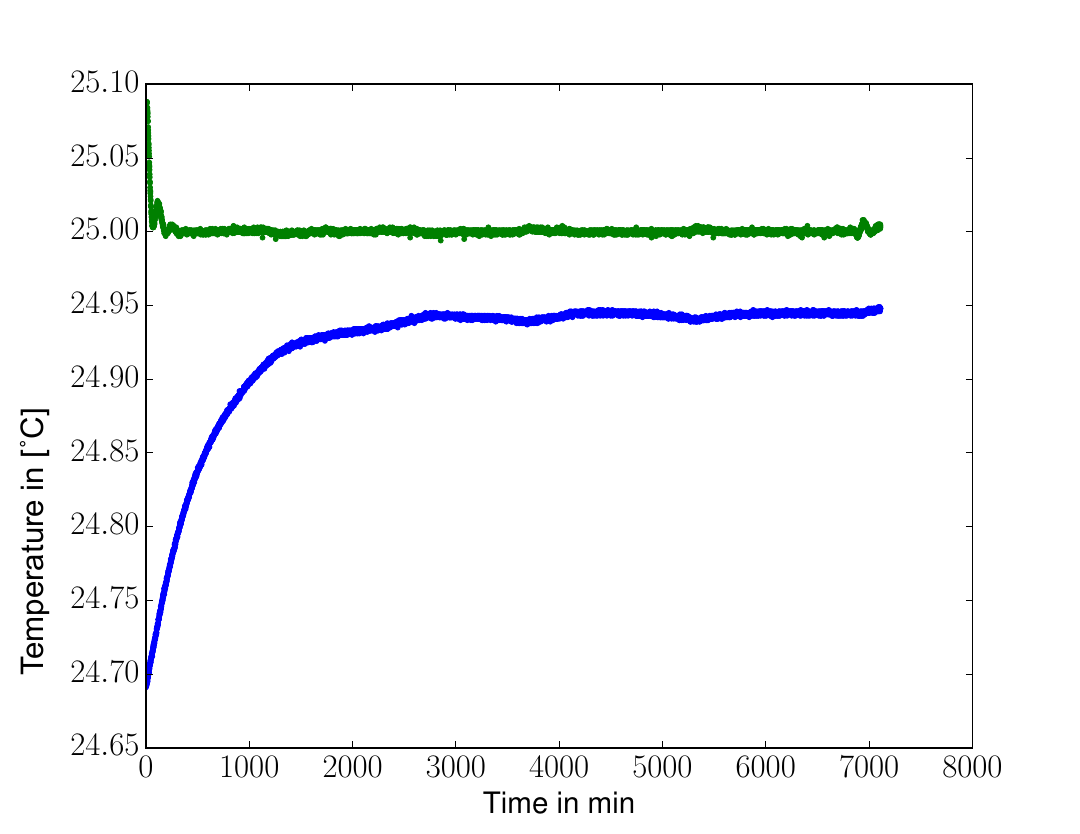}
\caption{Temperature of the RV-reference module after switching on the control loop. The green curve shows the control sensor located on the vacuum tank. The blue curve represents the sensor located in proximity of the \'etalon.}
\label{fig:21}
\end{figure}

We used a LakeShore 331 temperature controller with a set-point set at $25$\,C. The temperature control loop is closed on the sensor, which is bonded to the vacuum vessel's inside wall. A second sensor monitors the temperature of the opto-mechanical structure in proximity of the Fabry-P\'erot \'etalon. Figure \ref{fig:21} shows the temperature of both sensors after turning on the temperature controller.

The temperature of the vacuum tank stabilizes after less than one hour with a dispersion of $\sigma_{T}=1.2$\,mK, while the sensor close to the \'etalon reaches a stable temperature after about $40h$ with a slightly higher dispersion of $\sigma_{T}=2.1$\,mK. This can be explained by the fact that the test was performed in a laboratory with no temperature control at all and thus night-to-day temperature variations of several degrees. Even under these conditions the temperature stability of our RV-reference module is well within specification.

\section{Radial velocity performance}
\subsection{Photon noise}


By analogy with the HARPS-N Fabry-P\'erot we know that the centroid of a single emission line featuring 100000 $e^-$ per line can be determined with a photon-noise limited precision $\sigma$ of better than 2\,m\,s$^{-1}$. SPIRou will be using a Hawaii 4RG detector with lower dynamical range than typical CCD detector. If we assume a conservative line flux of only 25000 $e^-$ per exposure, given the fact that the present \'etalon will produce about 14280 lines in the spectral range of SPIROU, we will obtain a global precision of the drift measurement of
\begin{equation}
\sigma_{tot}=\frac{\sigma_{100000}\times \sqrt{100000/25000}}{\sqrt{14280}}=0.034 \,\, \textrm{m\,s}^{-1}.
\end{equation}

This value is significantly lower than the aimed precision of 0.3\,m\,s$^{-1}$ for SPIRou, and therefore compliant with the requirements.

\subsection{Radial velocity stability}
Radial-velocity stability measurements could however not be performed, because they would have required a spectral reference at least as stable as the RV-reference module is supposed to be. We therefore decided to postpone these measurements until they can be performed on SPIRou itself, where it will be possible to compare our module directly with a Uranium hollow-cathode lamp.\\

Nevertheless, we can refer to stability measurements of an identical system (besides the \'etalon coatings because they are optimized for the visible) made with HARPS-N \citep{Wildi2011}. Doing this, we demonstrated that the short-term RV-dispersion obtained using the Fabry-P\'erot system is better than that measured using the ThAr hollow-cathode lamp. Typical dispersion values are of the order of 0.1 to 0.2\,m\,s$^{-1}$ on time scales from hours to a day. This is again perfectly in line with the specified RV-stability requirement. Given the almost identical design of these two systems, we have plausible reason to expect similar performances on the SPIRou RV-reference module.

\subsection{Modal noise}
Several studies have demonstrated that the finite number of modes in a wave guide such as a fibres, may introduce noise in excess to pure photon noise. For instance, the modal noise limits the signal-to-noise ratio ($S/N$) \citep{McCoy11} measured in the continuum of the measured spectrum. This modal noise increases with wavelength because of the fewer modes in the fibre and it may therefore represent a limit to the NIR radial-velocity precision budget. We therefore analyze the situation for the SPIRou RV-reference module. The number of excited modes in a uniformly-illuminated fibre is inversely proportional to the wavelength and is given by the formula \citep{McCoy11}
\begin{equation}
M=\frac{1}{2}\left(\frac{\pi \cdot d \cdot NA}{\lambda}\right)^2,
\end{equation}
where $d$ is the fibre core diameter, $\lambda$ is the light’s wavelength, and $NA$ is the fibre's numerical aperture. In the wavelength domain of SPIRou and for the chosen fibre, we have $M=9941$ at $980\,nm$ and $M=1741$ at 2350\,nm. At first glance it appears that at short wavelengths there are enough modes for our Fabry-P\'erot system because we expect to obtain an $S/N$ at continuum of the order of the square root of the number of modes
\begin{equation}
\left(\frac{S}{N}\right)_{limit}\,\sim \, M^{0.784}
\end{equation}
\citep{Baudrand2001} meaning about 100 on the blue end of the spectrum. On the red end of the spectrum the mode-limited $S/N$ ratio decreases to about 30, which is even below what we expect to be the limit to reach a precision of 1\,m\,s$^{-1}$.\\
Let us take HARPS-N as the reference instrument. The Fabry-P\'erot spectrum recorded by HARPS-N includes approximately 20000 lines with average $S/N=187$ pixel$^{-1}$ and global photonic error on the drift measurement $\sigma_{RV}=0.044$\,m\,s$^{-1}$. In SPIRou the number of lines is approximately 14000. Assuming a similar S/N per line, and by analogy with HARPS-N, we estimate that the global photonic error on the drift measurement will be of the order of $\sigma_{RV}=0.062$\,m\,s$^{-1}$.

\begin{table}[h]
\centering 
\begin{tabular}{l l l l}
\hline\hline
\textbf{SPIRou} & \textbf{Lines} & \textbf{S/N} & \textbf{$\sigma_{RV}$}\\ 
\hline
& $14280$   & 187 & 0.062\,m\,s$^{-1}$\\ 
 &  & 100 & $\sim \, 0.12$\,m\,s$^{-1}$\\ 
 &  & 30 & $\sim \, 0.38$\,m\,s$^{-1}$ \\ 
\hline\hline
\end{tabular}
\caption{Measured $\sigma_{RV}$ for two limit cases of S/N.}
\label{tab:modalnoise}
\end{table}

Assuming an effective $S/N=100$ pixel$^{-1}$ and $S/N=30$ pixel$^{-1}$ at the two extreme ends of the spectral range of SPIRou, we computed the effective measurement error values (see Table \ref{tab:modalnoise}). Given these values, we see that the modal noise contribution to the error budget is  $0.38$\,m\,s$^{-1}$ even in the worst case, and for the average of the spectral range it is better than the $0.3$\,m\,s$^{-1}$ requirement of SPIRou.\\

The impact of modal-noise-reduced $S/N$ on the photo-centre measurement of a spectral line and thus its radial velocity is not yet fully understood because the impact depends on many factors including the optical design of the spectrograph. Nevertheless we think that this question deserves further tests. We have therefore started and are currently conducting laboratory tests, such as  mechanical agitation of the fibre, to understand the impact of modal noise in fibres in terms of radial-velocity measurement in general and on other systems, such as our RV-reference module in particular. The detailed analysis of modal noise lies however beyond the goals of this paper. We refer to Blind et al. (in prep.) for a detailed discussion and presentation of these modal-noise tests and their results, and point towards \citep{Micheau2012} and \citep{Micheau2015} for the specific case of SPIRou, for which several solutions to the modal-noise problem have been proposed.

\section{Conclusions}
In this paper we present the design of a new infrared Fabry-P\'erot RV-reference module for SPIRou and we describe its key properties and spectral characteristics in detail. The primary objectives were to build, test and finally operate a Fabry-P\'erot-based RV-reference module able to provide the needed spectral information over the full wavelength range of SPIRou to measure instrumental drift to a precision of 30\,cm\,$s^{-1}$. It must be remarked that we followed a quite strict top-down approach: 1) After establishing the requirements set by SPIRou we 2) converted them into requirements for the RV-referece module, 3) verified the assumption by analysis and simulations, 4) designed and manufactured the module according to the choices we made, and finally 5) tested the key performances in the laboratory. The obtained performances, at least those which could be verified in the laboratory, are compliant with the requirements. Only the RV-stability requirements had to be postponed to a later stage when the SPIRou spectrograph will be ready for system tests. Nevertheless, we know by analogy with the HARPS-N RV-reference module, that we can expect an RV-stability well below 1\,m\,$s^{-1}$. In summary, we demonstrate that the SPIRou reference module can replace any cathode spectral lamp or absorption cell for drift measurements at relatively moderate costs.\\

Our module also provides equally-spaced spectral lines of uniform intensity over an extremely wide spectral range, which could greatly help in measuring the detector geometry. Although the \'etalon does not have the intrinsic accuracy provided by an atomic transition and the mirror coatings may introduce slow variations of the phase (and thus cause group-delay dispersion, as per \citet{Wildi2011}), it should be remembered that the line separation and position is perfectly smooth and continuous in wavelength. This information can be used to improve the wavelength solution, in particular to determine the higher-order terms of the polynomials used to describe the pixel-to-wavelength relationship, which are badly constrained by the sparse lines of hollow-cathod lamps only. In the absence of commercial and affordable laser-frequency combs that cover the entire spectral range, we are currently investigating and implementing a combined use of hollow-cathode lamps and Fabry-P\'erot calibration. The former provides the accuracy (or global wavelength 'zero point'), whereas the latter provides the detector geometry and the pixel-to-pixel and intra-pixel responses. In fact, we are also considering the possibility of varying the line position of the Fabry-P\'erot by changing the (low) pressure inside the vacuum tank, to 'scan' and characterize the response of every detector pixel individually.\\

We can conclude that the presented RV-reference module for SPIRou is able to measure instrumental drifts at the required precision level and can replace the performance-limited hollow-cathode lamps and absorption cells with respect to this goal. Furthermore, we identify possible developments of the module and significant potential to convert it into a calibration source, which could turn the Fabry-P\'erot-based light source into a cost-effective alternative to laser-frequency combs, at least in the short and mid term.



\begin{acknowledgements} 
This work was supported by the Swiss
National Science Foundation (SNSF) grants nr. 200021-140649 200020-152721 and 200020-166227. The authors acknowledge his financial support.
\end{acknowledgements}



\bibliographystyle{aa} 
\bibliography{biblio.bib} 

\begin{thebibliography}{17}
\expandafter\ifx\csname natexlab\endcsname\relax\def\natexlab#1{#1}\fi

\bibitem[{Ball(2006)}]{ball2006}
Ball, D.~W. 2006, Field guide to spectroscopy, SPIE field guides (Bellingham
  (Wash.): SPIE Press), sPIE : The international society for optical
  engineering

\bibitem[{{Barnes} {et~al.}(2011){Barnes}, {Jeffers}, \& {Jones}}]{Barnes11}
{Barnes}, J.~R., {Jeffers}, S.~V., \& {Jones}, H.~R.~A. 2011, \mnras, 412, 1599

\bibitem[{{Baudrand} \& {Walker}(2001)}]{Baudrand2001}
{Baudrand}, J. \& {Walker}, G.~A.~H. 2001, \pasp, 113, 851

\bibitem[{{Chazelas} {et~al.}(2012){Chazelas}, {Pepe}, \&
  {Wildi}}]{Chazelas2012}
{Chazelas}, B., {Pepe}, F., \& {Wildi}, F. 2012, in \procspie, Vol. 8450,
  Modern Technologies in Space- and Ground-based Telescopes and Instrumentation
  II, 845013

\bibitem[{{Chazelas} {et~al.}(2010){Chazelas}, {Pepe}, {Wildi}, {Bouchy},
  {Perruchot}, \& {Avila}}]{Chazelas2010}
{Chazelas}, B., {Pepe}, F., {Wildi}, F., {et~al.} 2010, in \procspie, Vol.
  7739, Modern Technologies in Space- and Ground-based Telescopes and
  Instrumentation, 773947

\bibitem[{{Delfosse} {et~al.}(2013){Delfosse}, {Donati}, {Kouach},
  {H{\'e}brard}, {Doyon}, {Artigau}, {Bouchy}, {Boisse}, {Brun}, {Hennebelle},
  {Widemann}, {Bouvier}, {Bonfils}, {Morin}, {Moutou}, {Pepe}, {Udry}, {do
  Nascimento}, {Alencar}, {Castilho}, {Martioli}, {Wang}, {Figueira}, \&
  {Santos}}]{Delfosse13}
{Delfosse}, X., {Donati}, J.-F., {Kouach}, D., {et~al.} 2013, in SF2A-2013:
  Proceedings of the Annual meeting of the French Society of Astronomy and
  Astrophysics, ed. L.~{Cambresy}, F.~{Martins}, E.~{Nuss}, \& A.~{Palacios},
  497--508

\bibitem[{{McCoy} \& {Ramsey}(2011)}]{McCoy11}
{McCoy}, K. \& {Ramsey}, L. 2011, in Advanced Maui Optical and Space
  Surveillance Technologies Conference, E69

\bibitem[{Micheau {et~al.}(2012)Micheau, Bouchy, Pepe, Chazelas, Kouach,
  Parès, Donati, Barrick, Rabou, Thibault, Saddlemyer, Perruchot, Delfosse,
  Striebig, Gallou, Loop, \& Pazder}]{Micheau2012}
Micheau, Y., Bouchy, F., Pepe, F., {et~al.} 2012, SPIRou @ CFHT: fiber links
  and pupil slicer

\bibitem[{{Micheau} {et~al.}(2015){Micheau}, {Bouy{\'e}}, {Parisot}, \&
  {Kouach}}]{Micheau2015}
{Micheau}, Y., {Bouy{\'e}}, M., {Parisot}, J., \& {Kouach}, D. 2015, in
  \procspie, Vol. 9605, Techniques and Instrumentation for Detection of
  Exoplanets VII, 96051Q

\bibitem[{{Pepe} {et~al.}(2004){Pepe}, {Mayor}, {Queloz}, \& {Udry}}]{Pepe2004}
{Pepe}, F., {Mayor}, M., {Queloz}, D., \& {Udry}, S. 2004, in IAU Symposium,
  Vol. 202, Planetary Systems in the Universe, ed. A.~{Penny}, 103

\bibitem[{{Reiners} {et~al.}(2014){Reiners}, {Banyal}, \&
  {Ulbrich}}]{Reiners2014}
{Reiners}, A., {Banyal}, R.~K., \& {Ulbrich}, R.~G. 2014, \aap, 569, A77

\bibitem[{{Reiners} {et~al.}(2010){Reiners}, {Bean}, {Huber}, {Dreizler},
  {Seifahrt}, \& {Czesla}}]{Reiners2010}
{Reiners}, A., {Bean}, J.~L., {Huber}, K.~F., {et~al.} 2010, \apj, 710, 432

\bibitem[{{Schwab} {et~al.}(2015){Schwab}, {St{\"u}rmer}, {Gurevich},
  {F{\"u}hrer}, {Lamoreaux}, {Walther}, \& {Quirrenbach}}]{Schwab2015}
{Schwab}, C., {St{\"u}rmer}, J., {Gurevich}, Y.~V., {et~al.} 2015, \pasp, 127,
  880

\bibitem[{{Udry} {et~al.}(2007){Udry}, {Bonfils}, {Delfosse}, {Forveille},
  {Mayor}, {Perrier}, {Bouchy}, {Lovis}, {Pepe}, {Queloz}, \&
  {Bertaux}}]{Udry2007}
{Udry}, S., {Bonfils}, X., {Delfosse}, X., {et~al.} 2007, \aap, 469, L43

\bibitem[{{Wildi} {et~al.}(2010){Wildi}, {Pepe}, {Chazelas}, {Lo Curto}, \&
  {Lovis}}]{Wildi2010}
{Wildi}, F., {Pepe}, F., {Chazelas}, B., {Lo Curto}, G., \& {Lovis}, C. 2010,
  in \procspie, Vol. 7735, Ground-based and Airborne Instrumentation for
  Astronomy III, 77354X

\bibitem[{{Wildi} {et~al.}(2011){Wildi}, {Pepe}, {Chazelas}, {Lo Curto}, \&
  {Lovis}}]{Wildi2011}
{Wildi}, F., {Pepe}, F., {Chazelas}, B., {Lo Curto}, G., \& {Lovis}, C. 2011,
  in \procspie, Vol. 8151, Techniques and Instrumentation for Detection of
  Exoplanets V, 81511F

\bibitem[{{Ycas} {et~al.}(2012){Ycas}, {Quinlan}, {Diddams}, {Osterman},
  {Mahadevan}, {Redman}, {Terrien}, {Ramsey}, {Bender}, {Botzer}, \&
  {Sigurdsson}}]{Ycas2012}
{Ycas}, G.~G., {Quinlan}, F., {Diddams}, S.~A., {et~al.} 2012, Optics Express,
  20, 6631

\end{thebibliography}


\end{document}